%% file: McNu96_paper_arxiv.tex
\documentclass[10pt,Proceedings,InsideFigs,SingleSpace]{ascelike}
\usepackage{amsmath}
\usepackage{graphicx}
\usepackage{subfigure}
\usepackage{epsfig}
\usepackage{array}
\makeatletter
\renewcommand{\@maketitle}{%
  \newpage
  \null
    \ifthenelse{\boolean{Journal}}
               {\vspace{0.00in}}
               {\ifthenelse{\boolean{NewProceedings}}
                 {\vspace{0.40in}}
                 {\vspace{0.40in}}}
    \centering
    {\large\bfseries\@title\par}%
     \vspace{1.0em}%
    {\normalsize\normalfont
      \begin{tabular}[t]{c}%
        \@author
      \end{tabular}\par}%
  \par
  \vspace{0.5em}}%
\makeatother
\usepackage{texdraw}
\include{txdps}
\begin{document}
\title{\parbox{4.40in}{{\small in \emph{Mechanics of Deformation and Flow of Particulate Materials},\\
        Eds. C. S. Chang, A. Misra, R. Y. Liang, and M. Babic, ASCE, New York, pp. 91-104}}\\[1.25in]
       {\normalsize DEFORMATION MEASURES FOR GRANULAR MATERIALS}}
\author{Matthew R. Kuhn,\thanks{Assoc. Prof., Dept. of Civ. Engrg.,
Multnomah School of Engrg., Univ. of Portland, 5000 N. Willamette Blvd.,
Portland, OR  97203. e-mail: kuhn$@$up.edu}
\ Member, ASCE}
\maketitle
%
%
\begin{abstract}
\small
The paper presents a micromechanical representation of deformation in 2D
granular materials.
The representation is a generalization 
of K. Bagi's work and is based upon the
void-cell approach of M. Satake.
The general representation applies to a material region partitioned
into polygonal subregions.
This representation possesses a certain consistency
that allows for a unique assignment of the contribution that
each contact displacement makes to the
average deformation of an assembly.
The paper addresses construction of the particle graph
and appropriate data structures
for use with the Discrete Element Method.
The approach is applied in a numerical simulation of a two-dimensional assembly
of disks.
The author presents results of the distributions of
deformation and particle-group rotation, with
a resolution of about a single particle diameter.
Deformation was very nonuniform, even at low strains.
Micro-bands, thin linear zones of intense rotation, were also observed.
\normalsize
\end{abstract}
%
\newcommand{\bsigma}{\boldsymbol{\sigma}}
\section{1. Introduction}
Deformation of a granular material produces movements of 
individual particles.
Although the particles themselves may deform,
this deformation is localized near 
contacts, and deformation of the aggregate material results
primarily from the shifting of particle centers.
Several methods have been used to measure and visualize 
the deformation that results 
from particle movements.
Plotting the particles' movement or velocity vectors is likely the
simplest method for 2D assemblies, and this method has been used to
infer complex deformation structures within granular materials.
\citeN{Cundall:1982a} used this technique with the Discrete Element
Method to discover the presence of velocity discontinuities in a 
2D assembly of disks.
These discontinuities were organized along apparent shear
surfaces during a simulated biaxial compression test,
but they began to appear at stress levels well below
the peak stress.
\citeN{Williams} refined the use
of velocity vectors by subtracting the
mean velocity field from
the velocities of individual particles.
They produced vector plots that showed the deviations by individual 
particles from the average velocity field during biaxial compression.
These plots reveal the development of ``circulation cells,'' groups
of several dozen particles that rotate as groups even though the
mean vorticity of the assembly is zero.
\par
In an early study to define the linkage between deformation and particle
movement,
\citeN{Rowe:1962a} considered
three different regular crystal-like
packings of equal sized spheres and studied the manner in which
simple deformation patterns would cause the particles to move within a single
unit cell of material.
The linkage between deformation and particle movement is, of course,
more complex with random packings.
\citeN{Bardet:1991a} developed a technique for estimating the
deformation in the vicinity of a single particle by using
regression analysis to find an affine displacement
field that would approximate movement of the particle and a few of its
neighbors.
\citeN{Kuhn:1996a} used a similar technique to measure local deformations
and deformation gradients.
\par
In developing a generalization of Rowe's stress-dilatancy theory,
\citeN{Horne:1965b} and \citeN{Oda:1975a} viewed deformation as 
a mechanism that occurs along chains (or ``solid paths'') of
particles.
Deformation of an assembly produces compression or elongation
of the chain by folding the branch vectors between pairs of adjacent
particles.
\citeN{Cundall:1982a} observed that particle sliding occurs primarily
alongside chains that are aligned in the direction of the major
principle strain increment.
This view of deformation along particle chains complements the
experimental observations of \citeN{Drescher:1972a} and \citeN{Oda:1982a},
who used photoelastic models of particle assemblies.
These experiments revealed the presence of ``force chains,'' along
which the major principle stress is borne by 
highly loaded particles in chains that are preferentially oriented
in the direction of the stress.
\par
\begin{figure}
  \epsfig{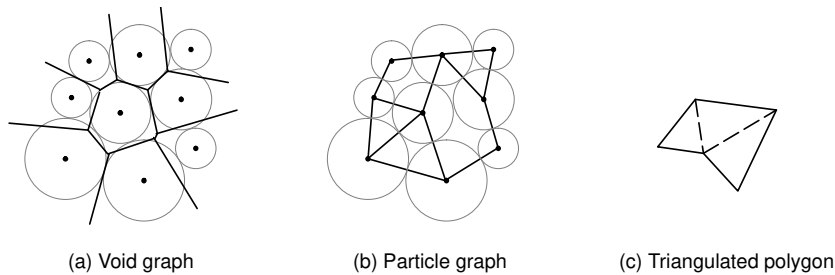}
  \caption{2D topological representations}
  \label{fig:voronoi3}
\end{figure}

In this paper, a more recent means of visualizing and measuring deformation
is pursued, one in which deformation occurs within the void space
between particles.
Such void-based methods require that an assembly be partitioned into
a covering of non-overlapping subregions, so that
the local effects of deforming the assembly can be measured within each
subregion.
\citeN{Bagi:1996a} classified a number of such partitioning schemes.
Among the simplest are those that use polygonal (2D) or
polyhedral (3D) subregions.
In one class of partitioning methods, the polygonal subregions encompass
individual particles, with the extremities of each polygon determined
by a preassigned rule (e.g., \citeNP{Annic:1993a}).
For example, in a Voronoi partition, all points within a polygonal
subregion are closer to a particular point (perhaps a particle's center)
than to other nearby points (Fig. \ref{fig:voronoi3}a).
Although efficient algorithms have been developed to produce Voronoi partitions
(see \citeNP{Orourke:1994a}),
an alternative partioning approach is used in this paper.
In the alternate class of partitions,
the corners of polygons are attached to material points
within the particles (usually particle centers), so that
the polygons represent \emph{void cells} (Fig. \ref{fig:voronoi3}b).
\citeN{Satake:1993a} introduced the concept of a \emph{particle graph}
and applied graph-theoretical methods to characterize displacements
of the particle centers and
develop duality relationships between the void and particle graphs
(Figs. \ref{fig:voronoi3}a and \ref{fig:voronoi3}b respectively).
His methods are the first of two bases for developments in this paper.
\par
Ostoja-Starzewski and his coworkers suggested the exclusive use of
triangular (simplex) subregions for 2D assemblies, requiring a triangulation of
the entire material region (see \citeNP{Ostoja:1995a}).
It should be noted, that any non-triangular polygon can be divided
into triangular subregions, and efficient algorithms have
been developed within the field of computational geometry to accomplish
this task (see Fig. \ref{fig:voronoi3}c and \citeNP{Orourke:1994a}).
\par
The second basis of this paper is the work of \citeN{Bagi:1996a}, who
derived an exact relationship between movements of
the vertices of triangular subregions
and the average deformation of an entire region.
In Section 2 a similar relationship is derived for polygonal
partitions.  
This is followed by an implementation with a dense
assembly of multi-sized disks.
Algorithms and data structures are presented in Section 3,
and results are presented in Section 4.
\section{2. Deformation of a Polygonally Partitioned Region}
This section considers a general closed 
2D material region $A$ which has been subdivided
into \small$\mathcal{L}$\normalsize~simple non-overlapping polygonal and closed subregions
$A^{i}$, each with a boundary $\partial A^{i}$, such that
\begin{equation} \label{eq:partition}
A \;=\; \underset{i}{\bigcup}\, A^{i}, \quad\quad
\underset{i}{\bigcap}\, A^{i} \;=\; 
\underset{i}{\bigcup}\, \partial A^{i} \;-\, \partial A \ , 
\quad\;
i \in \{1,2, \ldots , \mathcal{L}\}
\end{equation}
The region is represented by a planar graph of 
\small$\mathcal{L}$\normalsize~\emph{faces}
(``void cells'' in \citeNP{Satake:1993a}),
\small$\bar{\mathcal{M}}$\normalsize~\emph{edges} 
(``branches'' in \citeNP{Satake:1993a}),
and \small$\bar{\mathcal{N}}$\normalsize~\emph{vertices} 
(``nodes'' in \citeNP{Bagi:1996a}).
Instances of these three objects will be denoted by their respective
superscripts $i$, $j$, and~$k$.
Vertices are attached to material points within the region $A$.
\par
The spatial average of the velocity gradient $\mathbf{L}$
is denoted by $\bar{\mathbf{L}}$, 
with its Cartesian components defined by
\begin{equation} \label{eq:Ldef}
\bar{L}_{pq} \;\;\equiv\;\; \frac{1}{A} \int_A L_{pq} \, dA \;=\; 
\frac{1}{A} \int_A  v_{p,q} \, dA
\end{equation}
for velocity field $\mathbf{v}$.
Although the Gauss-Ostrogoski theorem can be applied to
evaluate $\bar{\mathbf{L}}$ by integrating along boundary $\partial A$,
this paper will concern an expression for 
$\bar{\mathbf{L}}$ in terms of the movements of
vertices and edges within the interior of $A$.
Toward this view, ($\ref{eq:partition}_1$) and (\ref{eq:Ldef})
are expressed as
\begin{equation} \label{eq:Lsum}
\bar{L}_{pq} 
\;\;=\;\; \frac{1}{A} \sum_{i} \int_{A^{i}} v_{p,q} \, dA
\;\;=\;\; \frac{1}{A} \sum_{i} A^{i} \bar{L}^{i}_{pq} \ , \quad
i \in \{1,2, \ldots , \mathcal{L}\} \ ,
\end{equation}
where $\bar{\mathbf{L}}^{i}$
is the average velocity gradient within the $i^{\mathrm{th}}$
polygonal face, $A^{i}$.
\par
\begin{figure}
\begin{center}
  \epsfig{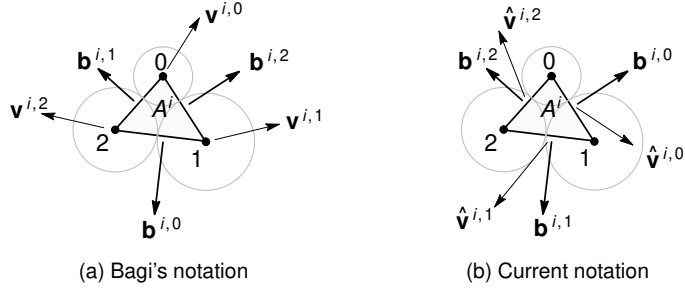}
  \caption{Triangular face: edge, velocity, and relative velocity vectors}
  \label{fig:triangle}
\end{center}
\end{figure}

\citeN{Bagi:1996a} derived an expression for $\bar{\mathbf{L}}^{i}$
of a single triangular region:
\begin{multline} \label{eq:Ltriangle}
\bar{L}^{i}_{pq} \;=\; 
\frac{1}{6A^{i}} 
\sum_{j_{1} < j_{2}} ( v_p^{i,k_{1}} - v_p^{i,k_{2}} ) \,
( b_q^{i,j_{1}} - b_q^{i,j_{2}} ), \\
j_{1}, \, j_{2} \in \{ 0,1,2 \}, \ k_{1}=j_{1}, \ k_{2}=j_{2} \ .
\end{multline}
In this expression, the three vertices of $A^{i}$ are locally labeled
with index $k$, so that $\mathbf{v}^{i,k}$ represents the velocity
of vertex $k$ of face $i$;
whereas, edge vector $\mathbf{b}^{i,j}$ corresponds to the
edge $j$ that lies opposite vertex $k(=j)$.
The magnitude of $\mathbf{b}^{i,j}$ is the length of
edge $j$, and its direction is the outward normal of that edge
(Fig. \ref{fig:triangle}a).
\par
Equation (\ref{eq:Ltriangle}) will be extended to a polygonal face $A^{i}$ with
$m^{i}$ edges (i.e., with \emph{valence} $m^{i}$).
The vertices and edges will be locally labeled $0$ through $m^{i} - 1$
in a consistent counterclockwise manner.
Departing from Bagi's notation, 
the label $j$ will be assigned to the edge 
connecting vertices $k_{1}=j$ and $k_{2}= j+1$ (Fig. \ref{fig:triangle}b).
Additions and subtractions that involve vertex or edge indices
will be implied modulo $m^{i}$, so that 
$\:j_{1} + j_{2} \mapsto j_{1} + j_{2}\! \pmod{m^{i}}$.
The relative velocity vector $\mathbf{\hat{v}}^{i,j}$ of edge $j$ is
defined as
\begin{equation} \label{eq:vhat_def}
\mathbf{\hat{v}}^{i,j} \;\;\equiv\;\; \mathbf{v}^{i,k_{2}(= j+1)} 
- \mathbf{v}^{i,k_{1}(= j)} \ .
\end{equation}
\citeN{Satake:1993a} and \citeN{Bagi:1996a}
presented separate relationships among
the $\mathbf{\hat{v}}^{i,j}$ and $\mathbf{b}^{i,j}$
vectors, 
\begin{equation} \label{eq:continuity}
\sum_{j=1}^{m^{i}} \mathbf{\hat{v}}^{i,j} \;=\; 0 \ , \quad\quad
\sum_{j=1}^{m^{i}} \mathbf{b}^{i,j} \;=\; 0 \ ,
\end{equation}
which will be used in the following developments.
Using the new notation, equations (\ref{eq:Ltriangle}), (\ref{eq:vhat_def}), 
and ($\ref{eq:continuity}_{1}$) are combined to give
\begin{equation} \label{eq:Q_with_triangle}
\bar{L}^{i}_{pq} \;\;=\;\;
\frac{1}{6A^{i}}
\sum_{j_{1},\, j_{2} \in\{0,1,\ldots,m-1\}}
Q^{m}_{j_{1}\: j_{2}}
\hat{v}_{p}^{i,\,j_{1}}\,
b_{q}^{i,\,j_{2}}
\end{equation}
for a triangular region, $m=3$.
Possible, but non-unique, forms of the $3 \times 3$ matrix
$\mathbf{Q}^{3}$ are
\begin{equation} \label{eq:Q_for_triangles}
\mathbf{Q}^{3} \;=\;
\left[
\begin{array}{rrr}
  0 &  1 & -1\\
 -1 &  0 &  1\\
\ 1 & -1 &  0\ 
\end{array}
\right]
\quad \text{or} \quad
\left[
\begin{array}{rrr}
 0 &\ 3 &\ 0 \\
-3 &  0 &  0 \\
 0 &  0 &  0
\end{array}
\right] \ .
\end{equation}
where the first of these matrices, ($\ref{eq:Q_for_triangles}_1$),
can also be written as
\begin{equation} \label{eq:Q_for_triangles2}
Q^{3}_{j_{1}\: j_{2}} \;=\; 
\begin{cases}
\ \ 0 & j_{2} - j_{1} = 0 \\
\ \ 1 & j_{2} - j_{1} = 1 \\
 -1 & j_{2} - j_{1} = 2
\end{cases}
\quad\quad
j_{1},\, j_{2} \in\{0,1,2\} \ ,
\end{equation}
with differences $j_{2} - j_{1}\! \pmod{3}$.
\par
The elements of
matrix $\mathbf{Q}^{m}$
will be derived for a general $m$-polygon.
The matrix is not unique, since the $\mathbf{\hat{v}}$ and $\mathbf{b}$
vectors of a polygon are constrained by ($\ref{eq:continuity}_1$)
and ($\ref{eq:continuity}_2$).
Uniqueness can be attained, however, by requiring that
$\mathbf{Q}^{m}$ exhibit a certain
\emph{consistency}, namely, that
\begin{equation} \label{eq:consistency}
q -p = s - r \;\;\;\Rightarrow\;\;\; Q^{m}_{pq} = Q^{m}_{rs} \ ,
\end{equation}
where the differences 
$q-p$ and $s-r$ (henceforth termed \emph{directed separations}) are 
computed modulo $m$.
Matrix ($\ref{eq:Q_for_triangles}_1$) satisfies this consistency
condition;
matrix ($\ref{eq:Q_for_triangles}_2$) does not.
Restriction (\ref{eq:consistency})
will lead to a unique $\mathbf{Q}^{m}$ matrix for
any two polygons of the same valence $m$.
The matrix can be readily incorporated into an algorithm for
computing local deformations within a region (Sections 3 and 4).
The consistent elements of $\mathbf{Q}^{m}$ are given recursively by
\begin{equation} \label{eq:Q_recursion}
Q^{m+1}_{pq} =
\begin{cases}
0 & q-p=0 \\
\frac{\displaystyle{1}}{\displaystyle{m+1}} (m Q^{m}_{pq} + 3) & q-p=1 \\
\frac{\displaystyle{1}}{\displaystyle{m+1}} 
[(m+1-\langle q-p \rangle ) Q^{m}_{pq} + 
\langle q-p \rangle Q^{m}_{p,q-1} ] & q-p \neq 0,1,m \\
\frac{\displaystyle{1}}{\displaystyle{m+1}} (m Q^{m}_{pq} - 3) & q-p=m
\end{cases}
\end{equation}
with differences $\langle q-p \rangle$, $q-1$, and $p-q$ 
all computed modulo $m$.
The seed matrix is
\begin{equation} \label{eq:Q_seed}
Q^{2}_{pq} \;=\; 0 \quad\quad p,q \in \{0,1\} \ .
\end{equation}
Components of matrices $\mathbf{Q}^{3}$
through $\mathbf{Q}^{6}$
are shown in Table \ref{table:Q_values}.
\begin{table}
\caption{Contents of matrix $\mathbf{Q^{m}_{pq}}$}
\label{table:Q_values}
\begin{center}
\small
\renewcommand{\arraystretch}{1.25}
\begin{tabular}{>{\small}c| >{\small}c >{\small}r >{\small}r >{\small}r>{\small}r  >{\small}r >{\small}r}
\hline\hline
&\multicolumn{7}{c}{$q - p \pmod{m}$} \\
\cline{2-8}
$m$ & 0 & 1 & 2 & 3 & 4 & 5 & 6 \\
\hline
3 & 0 & $\frac{3}{3}$ & $-\frac{3}{3}$ &  0 &    &     &       \\
4 & 0 & $\frac{6}{4}$ &  0 & $-\frac{6}{4}$ & 0  &     &       \\
5 & 0 & $\frac{9}{5}$ & $\frac{3}{5}$ & $-\frac{3}{5}$ & $-\frac{3}{5}$ &   0 & \\
6 & 0 & $\frac{12}{6}$ & $\frac{6}{6}$ &  0 & $-\frac{6}{6}$ & $-\frac{12}{6}$ & $\ $0 \\
\hline\hline
\end{tabular}
\end{center}
\end{table}
\par
Math\-em\-atical induction is used to derive the second case in
(\ref{eq:Q_recursion}), and other cases can be
similarly derived.
The values of $Q^{m}_{pq}$ for a triangle, $m=3$, 
are shown in first row of Table
\ref{table:Q_values} and correspond to the elements in
matrix ($\ref{eq:Q_for_triangles}_1$). 
These values are correctly given by
(\ref{eq:Q_recursion}).
The components of $Q^{m+1}_{pq}$ for a general $m\!+\!1$-polygon $A^{m+1}$
can be induced from those of an $m$-polygon $A^{m}$ 
by adding
triangle $A^{3}$ to form the extra vertex in $A^{m+1}$.
This process is illustrated in Figs. \ref{fig:poly2}a and \ref{fig:poly2}b,
where $\nu^{i,k}$ denotes
vertex $k$ of face $i$, and $e^{i,j}$ represents edge $j$ of
face $i$.
\begin{figure}
\begin{center}
  \epsfig{file=poly2.eps,width=5.00in,clip=}
  \caption{Combining $\mathbf{Q^{3}}$ and $\mathbf{Q^{m}}$ to form $\mathbf{Q^{m+1}}$ }
  \label{fig:poly2}
\end{center}
\end{figure}
\par
The average velocity gradient of the
$m\!+\!1$-polygon is the weighted sum of its two parents,
\begin{multline} \label{eq:Q_append}
\bar{\mathbf{L}}^{m+1} =  \\
- \frac{1}{6 A^{m+1}}
\left(\,
\sum_{p,q \in \{0,1,2\}} Q^{3}_{pq}\, \mathbf{\hat{v}}^{3,\,p}
\,\mathbf{b}^{3,\,q}
\;+
\sum_{p,q \in \{0,1,\ldots ,m-1\}}
Q^{m}_{pq}\, \mathbf{\hat{v}}^{m,\,p} 
\,\mathbf{b}^{m,\,q}
\right)
\ ,
\end{multline}
where juxtaposed vectors $\mathbf{\hat{v}} \, \mathbf{b}$
denote denote their dyadic product.
A direct computation of (\ref{eq:Q_append}) would produce
a non-consistent matrix $\mathbf{Q}^{m+1}$ even though
$\mathbf{Q}^{3}$ and $\mathbf{Q}^{m}$ are both consistent
by inductive assumption.
Consistency can be restored by finding an average
of the components $Q^{m+1}_{pq}$ that would result from
forming the $m\!+\!1$-polygon $A^{m+1}$ by separately adding a triangle to
an $m$-polygon at each of the $m+1$ vertices
(Fig. \ref{fig:poly2}c).
For a given directed separation \mbox{$q-p\! \pmod{m+1}$}, 
the contributions from all $m+1$ combination of
such $m$-polygons and triangles must be determined.
For example, adding a triangle to form vertex $\nu^{m+1,2}$ of
$A^{m+1}$ moves edge $e^{m,\alpha}$ of an $m$-polygon
into the interior of $A^{m+1}$ and replaces it with
edges $e^{m+1,1}$ and $e^{m+1,2}$ on $\partial A^{m+1}$.
Noting that ($\ref{eq:continuity}_2$) requires
\begin{equation}
\mathbf{b}^{m,\alpha} \; ( = - \mathbf{b}^{3,\beta}) \;\;=\;\;
\mathbf{b}^{m+1,1} + \mathbf{b}^{m+1,2} \ ,
\end{equation}
the matrix element $Q^{m}_{\alpha -1,\alpha}$ of the
$m$-polygon is carried over to both $Q^{m+1}_{01}$ and $Q^{m+1}_{02}$,
which correspond to directed separations of 1 and~2.
The added triangle also contributes a value of 3 to
element $Q^{m+1}_{12}$ and $-3$ to element $Q^{m+1}_{21}$,
as suggested in ($\ref{eq:Q_for_triangles}_2$).
Likewise, adding a triangle to form $\nu^{m+1,1}$ would
deliver a value of $3$ to $Q^{m+1}_{01}$.
Adding a triangle to form any of the other vertices of $A^{m+1}$
would each contribute the value $Q^{m}_{\alpha -1,\alpha}$.
to $Q^{m+1}_{01}$.
The consistent average for $Q^{m+1}_{01}$ or for any 
element $Q^{m+1}_{pq}$ with
directed separation $1\ (=q-p)$ is, therefore,
\begin{equation} \label{eq:Q_sep_1}
Q^{m+1}_{pq} \;=\; \frac{1}{m + 1} ( m Q^{m}_{pq} + 3 ) \quad\quad
q - p\!\!\!\pmod{m + 1} = 1 \ ,
\end{equation}
which corresponds to the second case of (\ref{eq:Q_recursion}).
The other cases can be similarly derived.
\section{3. Implementation Issues}
In this section, the planar graph of Section~2 is applied
to a 2D granular material. A method for constructing 
the graph is also presented,
and compact data structures are suggested for its efficient use with
the Discrete Element Method (DEM).
\subsection{Planar graph of a 2D granular material}
\begin{figure}
\begin{center}
  \epsfig{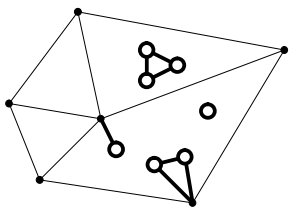}
  \caption{Pendant, island, peninsular, and isolated particles}
  \label{fig:pendant}
\end{center}
\end{figure}

As with the particle graph of \citeN{Satake:1993a}, vertices are
attached to particle centers,
edges correspond to lines that connect
contacting particle pairs, and faces $A^{i}$ represent voids.
Certain particles will be excluded from the graph---isolated 
and pendant particles, as well as island and peninsular 
particle groups---since
these particle are not part of the load bearing framework of the material
(Fig. \ref{fig:pendant}).
The number of vertices \small$\bar{\mathcal{N}}\:$\normalsize will, 
therefore, be less than
or equal to the number of particles \small$\mathcal{N}$\normalsize; 
likewise,
the number of edges \small$\bar{\mathcal{M}}\:$\normalsize will be no more than
the number of particle contacts \small$\mathcal{M}$\normalsize.
\subsection{Constructing the graph}
\begin{figure}
\begin{center}
  \epsfig{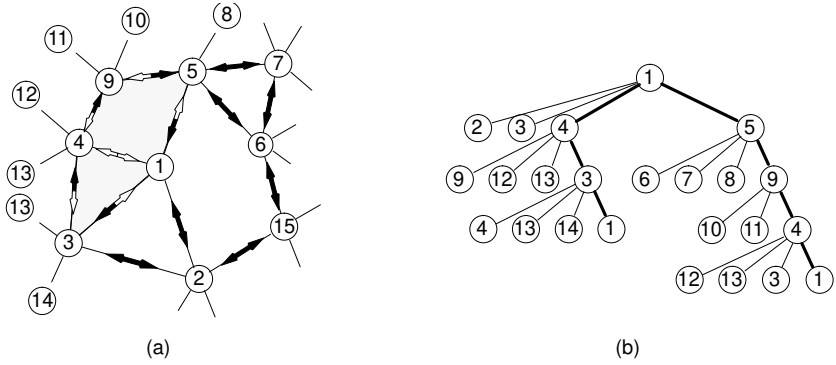}
  \caption{Identifying void-cell faces}
  \label{fig:graphs_build}
\end{center}
\end{figure}

The void-cell faces of the particle graph can be identified in
$\mathrm{O}(N)$ time.
The process is preconditioned by first constructing a doubly
connected linked list (DCLL) of all
edges (next subsection and \citeNP{Knuth:1973a}).
This list corresponds to assigning both
``coming and going'' directions to each contact in the assembly
(the solid arrows in Fig. \ref{fig:graphs_build}a).
The edges of each face are identified in a counterclockwise manner,
and once a face has been traversed, the corresponding edge
directions are removed from the DCLL and become ``inactive''.  
This removal prevents the same (or an overlapping)
face from being identified later.
\par
Faces are identified in the course of scanning each of the 
\small$\mathcal{N}\:$\normalsize particles.
When a particle's turn has arrived (such as particle number~1 in
Fig. \ref{fig:graphs_build}a), it becomes the root of a 
tree, which is then constructed one level at a time.
The tree's first level corresponds to all active
edges that depart from the root vertex.
Additional levels are added until a
limb (edge) completes a circuit that returns to the root vertex
(Fig. \ref{fig:graphs_build}b).
At each corner of a circuit, the next edge is chosen
to form the sharpest (clockwise) corner.
After completing the graph, it must be
scanned with the purpose of eliminating
pendants, islands, peninsular, and isolated vertices and edges
(Fig. \ref{fig:pendant}).
\subsection{Data structures}
A number of data structures are required for DEM implementation,
with each structure corresponding to a separate computation process.
These processes and their frequency are 
listed in Table \ref{table:Data_structures}.
\begin{table}
\caption{Implementation processes with the Discrete Element Method}
\label{table:Data_structures}
\begin{center}
\small
\begin{tabular}{
>{\small}l| >{\small}c|>{\small}c
}
\hline\hline
        & Data      & Frequency, \\
Process & structure & time steps \\
\hline
Near-neighbor identification/storage & SLL$_1$ & $\times 10^2$ \\
Contact detection                    &  ---    & $1$ \\
Contact data storage                 & SLL$_2$ & $1^u$ \\
Graph construction                   & DLL   & $1$ or $1^u$ \\
Graph storage                        & DCEL  & $1$ or $1^u$ \\
\hline
\multicolumn{3}{l}{\footnotesize $^u$ updating only} \\
\hline\hline
\end{tabular}
\end{center}
\end{table}
In one process, a complete list of \emph{near neighbors} is
assembled, where two particles are considered near neighbors if
they are in contact or are separated by no more than some maximum distance
(e.g., a proportion of $D_{50}$, Section 4).
The near neighbors are stored in a singly (linear) linked list
(SLL, see \citeNP{Knuth:1973a}).
Near neighbor detection is a lengthy process, but it need not be
done in each DEM time step.
With dense assemblies and an ample maximum separation criteria,
near neighbor detection might only be required every several hundred
time steps.
The contact detection process, however, must be performed each time step
but with the candidate contacts taken only from the near neighbor list.
A large quantity of data may need to be stored for each contact
(e.g., yield surface descriptors, hardening moduli, etc.).
To avoid wasting data storage space on non-contacting near neighbors,
a list of links is maintained between the near neighbor SLL and
the shorter arrays of contact data.
\par
Graph construction requires the use of a doubly linked list
(DLL, \citeNP{Knuth:1973a}), as was previously described.  
This DLL can be quickly constructed from the SLL of contacts.
The topologic relationships among vertices, edges, and faces are
efficiently stored and retrieved with a doubly connected
edge list (DCEL, \citeNP{Preparata:1985a}).
For the simulations described in Section 4, the graph was freshly
constructed at every time step, although it is possible that,
once constructed, its DCEL need only be updated to account for the
addition and loss of individual contacts.
\section{4. DEM Implementation and Results}
\begin{figure}
\centering
\mbox{
 \subfigure[\sffamily Particles]
    {\input{Test8_i_cell_def_circles}}
 \quad
 \subfigure[\sffamily Particle graph]
    {\input{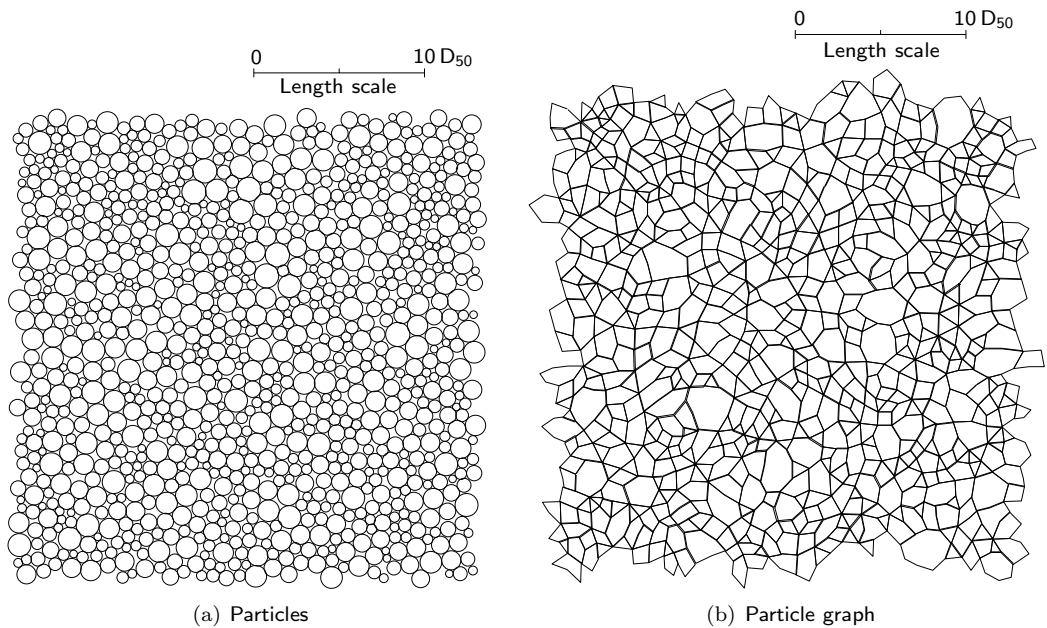}}
}
\caption{Assembly of 1002 particles}
\label{fig:assembly}
\end{figure}

Tests were performed on an assembly of 1002 irregularly arranged circular
disks by using the Discrete Element Method (Fig. \ref{fig:assembly}a).
The distribution of particle sizes was fairly uniform,
with the largest and smallest having
diameters $1.4$ and $0.45$ times the weight-median diameter $D_{50}$.
The coefficient of uniformity, $D_{60}/D_{10}$, was $1.7$.
The particle arrangement was initially random, isotropic, and
homogeneous, and the contact indentations
were small---on average, less than $0.02\%$ of $D_{50}$.
The initial void ratio was $0.189$.
The particle graph initially included $762$ faces 
(\small$\mathcal{L}$\normalsize),
$1633$ edges (\small$\bar{\mathcal{M}}$\normalsize),
and $871$ vertices (\small$\bar{\mathcal{N}}$\normalsize), 
and it is shown in Fig. \ref{fig:assembly}b.
The width of an average face was about one particle diameter ($D_{50}$).
The initial void ratio was $0.189$.
The low void ratio and an effective coordination number 
(\small$2\bar{\mathcal{M}} / \bar{\mathcal{N}}$\normalsize) of $3.75$
correspond to a fairly dense assembly \cite{Rothenburg:1992a}.
The assembly was surrounded by two pairs of periodic boundaries, 
which bestow a long-range translational symmetry in both
the vertical and horizontal directions.
The resulting graph is homeomorphic with a torus, for which the
Euler formula is
\begin{equation} \label{eq:Euler}
\mathcal{L} - \bar{\mathcal{M}} + \bar{\mathcal{N}} \;\;=\;\; 0 \ .
\end{equation}
\par
The contact force mechanism between particles consisted of
normal and tangential (linear) springs of
the same stiffness.  
Contact sliding would occur abruptly when the friction coefficient of
$0.50$ was attained.
\par
\begin{figure}
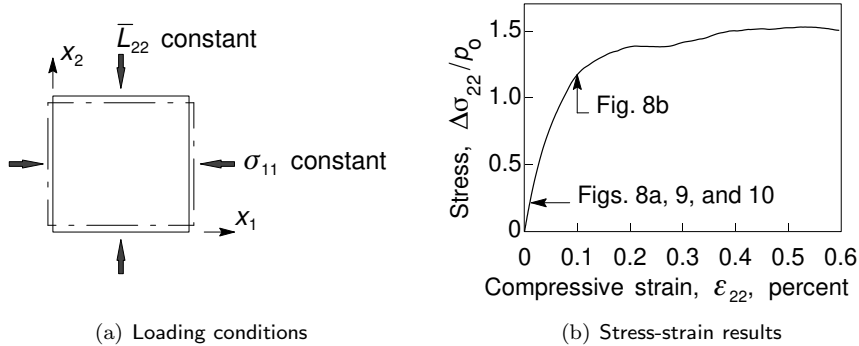

\centering
 \subfigure[\sffamily Loading conditions]
  {\epsfig{file=loading.eps}
  }
 \quad
 \subfigure[\sffamily Stress-strain results]
  {\epsfig{file=crs.eps}
  }
\caption{Biaxial compression test}
\label{fig:stress_strain}
\end{figure}

The assembly was loaded by moving the upper and lower boundaries toward each
other at a constant rate (Fig. \ref{fig:stress_strain}a).
The average stress $\mathbf{\bar{\bsigma}}$ within the assembly was monitored
by averaging the contributions from
each edge:
\begin{equation} \label{eq:stress_average}
\bar{\sigma}_{pq} \;\;=\;\; \sum_{j=1}^{\bar{\mathcal{M}}}
f^{j}_{p} \, d^{j}_{q}
\end{equation}
(\citeNP{Christoffersen:1981a}, \emph{cf.\ }\citeNP{Love:1927a} pp~618-9),
where $\mathbf{d}^{j}$ is the edge vector and $\mathbf{f}^{j}$
is the contact force of edge $j$.
The distance between the vertical sides was continually adjusted
to maintain constant horizontal stress $\bar{\sigma}_{11}$.
The measured stress-strain behavior is plotted 
in Fig. \ref{fig:stress_strain}b.
\par
\begin{figure}
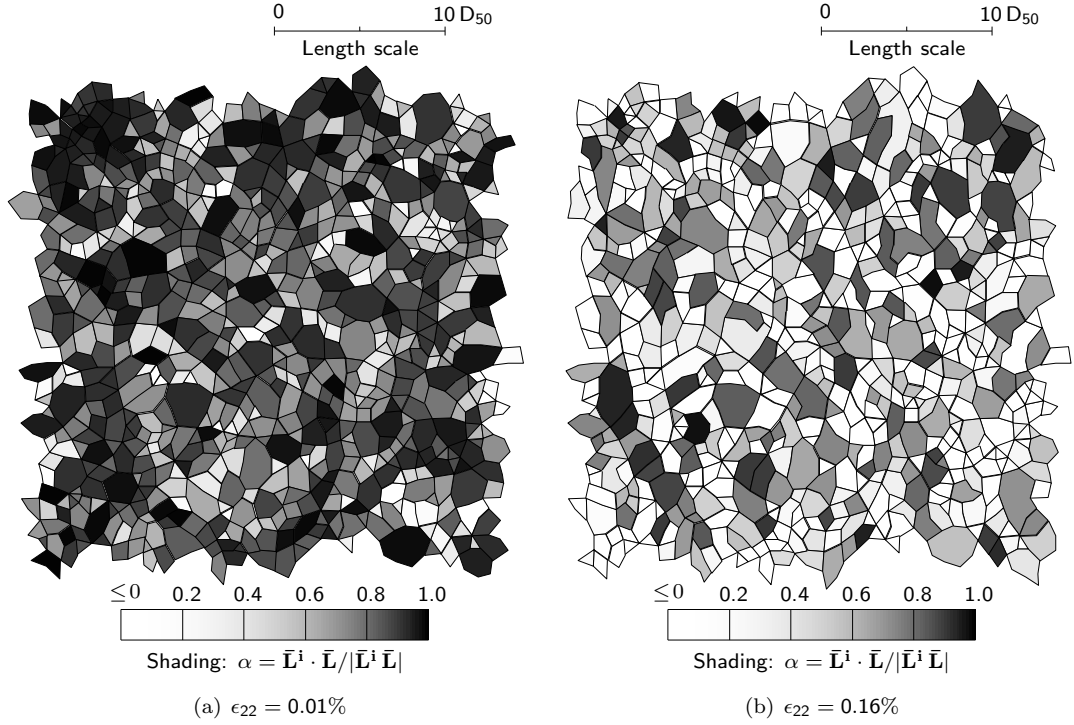

\centering
\mbox{
 \subfigure[\sffamily \rule{0mm}{8mm}$\mathsf{\epsilon_{22} = 0.01\%}$]
    {\input{Test8a1_cell_def_1a}}
 \quad
 \subfigure[\sffamily \rule{0mm}{8mm}$\mathsf{\epsilon_{22} = 0.16\%}$]
    {\input{Test8a2_cell_def_1a}}
}
\caption{Alignment of local deformations $\mathbf{\bar{L}^{i}}$
         with the average deformation $\mathbf{\bar{L}}$}
\label{fig:alignment}
\end{figure}

The assembly's average velocity gradient $\bar{\mathbf{L}}$ could be trivially
computed from the motions of the boundaries 
rather than by averaging the gradients $\bar{\mathbf{L}}^{i}$
within the void-cell faces.
Expression (\ref{eq:Q_with_triangle}) for $\bar{\mathbf{L}}^{i}$
was instead used to investigate the distribution of
deformation within the assembly.
As a direction application, the uniformity of deformation
within the assembly was investigated on the microscale of a 
single void-cell.
These measurements serve to test the assumption of
uniform deformation that has been made by several investigators
to estimate assembly stiffness on the basis of contact stiffness
characteristics.
One measure of deformation uniformity is the \emph{alignment}
of the local and average velocity gradients,
$\mathbf{\bar{L}}^{i}$ and $\mathbf{\bar{L}}$.
Alignment $\alpha$ is defined as the direction cosine
between the two tensors:
\begin{equation}
\alpha \;\;=\;\; \frac{\mathbf{\bar{L}}^{i} \mathbf{\cdot} \mathbf{\bar{L}}}
{|\mathbf{\bar{L}}^{i}| \, |\mathbf{\bar{L}}|} \ ,
\end{equation}
where an appropriate inner product and its associated norm are
\begin{equation}
\mathbf{\bar{L}}^{i} \mathbf{\cdot} \mathbf{\bar{L}} \;=\;
\bar{L}_{pq}^{i} \bar{L}_{pq} \quad \text{and} \quad
|\mathbf{\bar{L}}^{i}| \;=\; 
\left( \bar{L}_{pq}^{i} \bar{L}_{pq}^{i}\right)^{1/2}\ .
\end{equation}
Figures \ref{fig:alignment}a and \ref{fig:alignment}b show the distribution of
$\alpha$ when the strain is $0.01\%$ and $0.16\%$.
Even at the lower strain, when material behavior is nearly elastic,
deformation is quite variable, with an average
$\alpha$ of $0.73$.
At the larger strain, where stress $\Delta \sigma_{22}$ is within $14\%$
of its peak value, the deformation is extremely nonuniform.
About $22\%$ of void cells have an $\alpha$ less than
or equal to zero (i.e., white shading), which corresponds
to local deformation occuring in a direction opposite the average deformation.
\par
\begin{figure}
\begin{center}
  \input{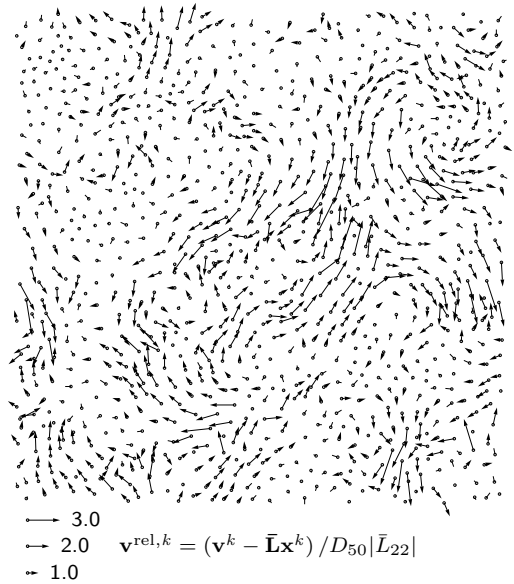}
  \caption{Particle relative movement vectors at
           $\mathbf{\boldsymbol{\epsilon}_{22} = 0.01\%}$}
  \label{fig:movements}
\end{center}
\end{figure}

Another interesting result is the rotaion of void-cell faces.
Figures \ref{fig:rotation}a and \ref{fig:rotation}b show the distribution of
\begin{figure}
\centering
\mbox{
 \subfigure[\sffamily \rule{0mm}{8mm}Clockwise]
    {\input{Test8a1_cell_def_3b}}
 \quad
 \subfigure[\sffamily \rule{0mm}{8mm}Counterclockwise]
    {\input{Test8a1_cell_def_3a}}
}
\caption{Local vorticity at $\mathbf{\boldsymbol{\epsilon}_{22} = 0.01\%}$}
\label{fig:rotation}
\end{figure}
normalized vorticity $w$, defined as
\begin{equation}
w \;\;=\;\; -\frac{\bar{L}_{12}^{i} - \bar{L}_{21}^{i}}
{|\mathbf{\bar{L}}|} \ .
\end{equation}
The two figures show clockwise and counterclockwise values
with a gray scale that accentuates direction but deemphasizes
magnitude.
These figures can be compared with Fig. \ref{fig:movements}
that shows relative velocity vectors $\mathbf{v}^{\mathrm{rel}}$,
defined in a similar manner as in \cite{Williams}, but
with nondimensional form
\begin{equation}
\mathbf{v}^{\mathrm{rel},\,k} \;\;=\;\;
\frac{\mathbf{v}^{k} - \mathbf{\bar{L}} \mathbf{x}^{k}}
{D_{50}\, |\bar{L}_{22}|} \ .
\end{equation}
Circulation cells \cite{Williams} are apparent in
Figs.~\ref{fig:rotation} and~\ref{fig:movements},
but the distributions of vorticity in
Fig.~\ref{fig:rotation} reveal a structure to the
circulations.
Intense rotations are frequently organized in thin bands that
are inclined relative to the principle stretch directions.
These zones, termed \emph{micro-bands}, are as thin as one or two
void-cell widths.
Such micro-bands may play an important role in the deformation
of granular materials.
\section{Conclusions}
A method is presented for determining localized
deformation within a polygonally partitioned 2D region.
The method was applied to an assembly of multi-sized disks by using
the Discrete Element Method.
Deformation was very nonuniform, even at low strains.
Micro-bands, thin zones of intense rotation, were also observed.
The method may be useful in revealing other deformation structures
and lead to realistic constitutive descriptions that are
based upon the micromechanics of granular materials.
\setlength{\bibindent}{1em}
\small

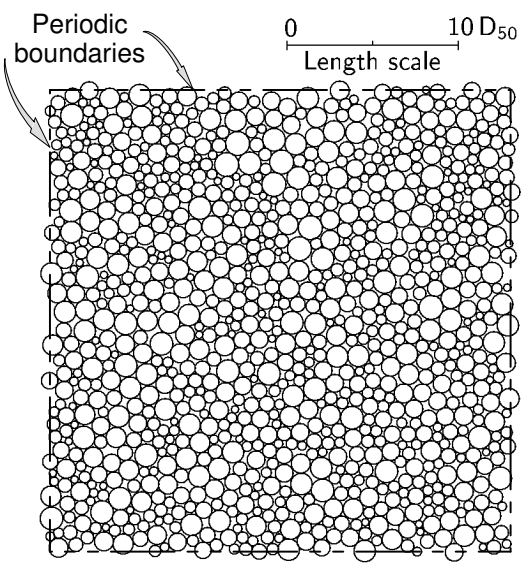
\end{document}

%% file: Test8_i_cell_def_circles.tex
\footnotesize\sffamily
\begin{texdraw}
\drawdim in
\setunitscale 0.45
\linewd 0.004
\PSsetlinecap 0
\def\Btext #1{\bsegment
\textref h:C v:T \htext (0 -0.10){#1}
\esegment}
\def\Ttext #1{\bsegment
\textref h:C v:B \htext (0 +0.07){#1}
\esegment}
\def\Tmtext #1{\bsegment
\textref h:C v:B \htext (-0.06 +0.04){#1}
\esegment}
\def\Rtext #1{\bsegment
\textref h:L v:C \htext (-0.04 0){#1}
\esegment}
\move(+3.219 +3.221)
\lcir r:+0.056
\move(+4.539 +2.293)
\lcir r:+0.080
\move(+4.449 +4.135)
\lcir r:+0.093
\move(+4.988 +1.333)
\lcir r:+0.133
\move(+4.498 +1.451)
\lcir r:+0.054
\move(+2.274 +0.038)
\lcir r:+0.071
\move(+1.381 +3.397)
\lcir r:+0.045
\move(+1.864 +1.976)
\lcir r:+0.098
\move(+4.495 +1.758)
\lcir r:+0.101
\move(+3.126 +0.239)
\lcir r:+0.090
\move(+1.415 +4.505)
\lcir r:+0.096
\move(+4.052 +2.474)
\lcir r:+0.065
\move(+0.229 +0.404)
\lcir r:+0.070
\move(+0.189 +0.972)
\lcir r:+0.112
\move(+1.026 +3.985)
\lcir r:+0.071
\move(+1.236 +5.116)
\lcir r:+0.051
\move(+3.861 +5.340)
\lcir r:+0.093
\move(+0.858 +4.047)
\lcir r:+0.107
\move(+0.689 +3.011)
\lcir r:+0.060
\move(+3.520 +1.325)
\lcir r:+0.102
\move(+4.916 +0.390)
\lcir r:+0.100
\move(+2.497 +4.019)
\lcir r:+0.074
\move(+4.849 +4.783)
\lcir r:+0.058
\move(+0.863 +5.140)
\lcir r:+0.097
\move(+0.572 +0.906)
\lcir r:+0.093
\move(+4.070 +1.049)
\lcir r:+0.065
\move(+4.284 +0.816)
\lcir r:+0.076
\move(+1.424 +4.674)
\lcir r:+0.074
\move(+2.379 +0.660)
\lcir r:+0.054
\move(+3.087 +3.584)
\lcir r:+0.063
\move(+5.128 +3.241)
\lcir r:+0.123
\move(+4.579 +1.263)
\lcir r:+0.059
\move(+3.228 +0.352)
\lcir r:+0.063
\move(+4.227 +0.987)
\lcir r:+0.104
\move(+2.341 +1.714)
\lcir r:+0.092
\move(+3.976 +5.151)
\lcir r:+0.129
\move(+2.409 +0.475)
\lcir r:+0.054
\move(+3.360 +5.365)
\lcir r:+0.110
\move(+4.636 +3.797)
\lcir r:+0.084
\move(+1.964 +4.727)
\lcir r:+0.081
\move(+2.520 +1.724)
\lcir r:+0.087
\move(+5.042 +0.218)
\lcir r:+0.113
\move(+5.065 +4.272)
\lcir r:+0.045
\move(+0.754 +2.297)
\lcir r:+0.047
\move(+4.161 +2.064)
\lcir r:+0.116
\move(+2.232 +3.875)
\lcir r:+0.107
\move(+2.599 +4.274)
\lcir r:+0.138
\move(+3.028 +4.629)
\lcir r:+0.078
\move(+4.989 +3.773)
\lcir r:+0.096
\move(+4.587 +1.578)
\lcir r:+0.101
\move(+1.199 +3.398)
\lcir r:+0.053
\move(+2.231 +4.168)
\lcir r:+0.087
\move(+0.693 +5.273)
\lcir r:+0.119
\move(+4.804 +3.750)
\lcir r:+0.091
\move(+2.603 +1.601)
\lcir r:+0.062
\move(+3.580 +2.282)
\lcir r:+0.125
\move(+2.364 +3.548)
\lcir r:+0.085
\move(+2.078 +2.697)
\lcir r:+0.046
\move(+3.813 +0.311)
\lcir r:+0.113
\move(+3.010 +2.076)
\lcir r:+0.058
\move(+1.303 +2.110)
\lcir r:+0.056
\move(+4.855 +2.005)
\lcir r:+0.072
\move(+2.792 +0.266)
\lcir r:+0.116
\move(+3.452 +2.891)
\lcir r:+0.095
\move(+1.791 +1.079)
\lcir r:+0.080
\move(+1.773 +2.137)
\lcir r:+0.087
\move(+0.440 +0.171)
\lcir r:+0.077
\move(+1.113 +3.868)
\lcir r:+0.075
\move(+2.970 +0.290)
\lcir r:+0.063
\move(+1.336 +0.061)
\lcir r:+0.051
\move(+4.890 +3.623)
\lcir r:+0.062
\move(+4.255 +1.250)
\lcir r:+0.062
\move(+3.251 +2.832)
\lcir r:+0.115
\move(+4.088 +1.248)
\lcir r:+0.049
\move(+3.864 +4.440)
\lcir r:+0.070
\move(+2.390 +4.231)
\lcir r:+0.075
\move(+3.378 +0.682)
\lcir r:+0.081
\move(+1.886 +2.916)
\lcir r:+0.076
\move(+5.019 +0.060)
\lcir r:+0.047
\move(+2.125 +1.161)
\lcir r:+0.125
\move(+0.105 +0.050)
\lcir r:+0.113
\move(+4.658 +3.195)
\lcir r:+0.126
\move(+4.859 +0.541)
\lcir r:+0.062
\move(+3.427 +5.205)
\lcir r:+0.063
\move(+2.417 +4.360)
\lcir r:+0.054
\move(+2.205 +0.974)
\lcir r:+0.079
\move(+2.065 +3.860)
\lcir r:+0.061
\move(+4.248 +4.608)
\lcir r:+0.069
\move(+5.087 +4.589)
\lcir r:+0.121
\move(+0.326 +2.079)
\lcir r:+0.049
\move(+1.733 +2.357)
\lcir r:+0.120
\move(+2.650 +3.492)
\lcir r:+0.046
\move(+0.389 +3.606)
\lcir r:+0.051
\move(+1.903 +2.247)
\lcir r:+0.083
\move(+0.242 +3.532)
\lcir r:+0.060
\move(+0.170 +2.071)
\lcir r:+0.088
\move(+3.970 +1.374)
\lcir r:+0.099
\move(+1.055 +2.895)
\lcir r:+0.097
\move(+4.609 +5.042)
\lcir r:+0.101
\move(+3.785 +1.405)
\lcir r:+0.064
\move(+2.175 +3.226)
\lcir r:+0.100
\move(+0.265 +5.301)
\lcir r:+0.086
\move(+1.683 +4.492)
\lcir r:+0.051
\move(+1.208 +0.941)
\lcir r:+0.055
\move(+4.228 +2.680)
\lcir r:+0.081
\move(+0.594 +1.465)
\lcir r:+0.106
\move(+0.306 +0.091)
\lcir r:+0.078
\move(+4.998 +2.096)
\lcir r:+0.098
\move(+3.088 +4.323)
\lcir r:+0.074
\move(+5.226 +0.857)
\lcir r:+0.134
\move(+1.995 +2.596)
\lcir r:+0.084
\move(+3.300 +4.892)
\lcir r:+0.128
\move(+3.653 +4.345)
\lcir r:+0.054
\move(+4.625 +1.119)
\lcir r:+0.092
\move(+1.421 +2.092)
\lcir r:+0.061
\move(+1.790 +3.158)
\lcir r:+0.120
\move(+1.623 +1.864)
\lcir r:+0.067
\move(+5.129 +4.387)
\lcir r:+0.085
\move(+3.261 +4.054)
\lcir r:+0.085
\move(+5.089 +5.280)
\lcir r:+0.077
\move(+2.447 +1.551)
\lcir r:+0.102
\move(+0.383 +5.222)
\lcir r:+0.055
\move(+1.508 +2.313)
\lcir r:+0.110
\move(+3.704 +2.133)
\lcir r:+0.062
\move(+4.515 +0.484)
\lcir r:+0.057
\move(+3.743 +2.571)
\lcir r:+0.068
\move(+2.458 +3.640)
\lcir r:+0.046
\move(+1.978 +3.995)
\lcir r:+0.098
\move(+4.545 +3.914)
\lcir r:+0.064
\move(+3.570 +2.796)
\lcir r:+0.056
\move(+2.037 +1.860)
\lcir r:+0.110
\move(+0.738 +4.968)
\lcir r:+0.116
\move(+1.448 +1.521)
\lcir r:+0.048
\move(+2.353 +3.404)
\lcir r:+0.053
\move(+2.702 +1.959)
\lcir r:+0.062
\move(+4.495 +1.342)
\lcir r:+0.056
\move(+0.360 +0.998)
\lcir r:+0.061
\move(+4.263 +0.011)
\lcir r:+0.053
\move(+0.192 +1.410)
\lcir r:+0.075
\move(+1.382 +2.669)
\lcir r:+0.127
\move(+2.172 +3.454)
\lcir r:+0.129
\move(+1.261 +0.636)
\lcir r:+0.080
\move(+2.327 +2.363)
\lcir r:+0.049
\move(+0.608 +2.892)
\lcir r:+0.060
\move(+1.306 +4.214)
\lcir r:+0.091
\move(+4.676 +1.733)
\lcir r:+0.077
\move(+4.096 +4.525)
\lcir r:+0.104
\move(+3.909 +2.564)
\lcir r:+0.098
\move(+4.322 +4.211)
\lcir r:+0.055
\move(+0.005 +5.127)
\lcir r:+0.056
\move(+1.966 +1.650)
\lcir r:+0.113
\move(+1.619 +1.482)
\lcir r:+0.128
\move(+1.688 +1.995)
\lcir r:+0.079
\move(+1.458 +0.289)
\lcir r:+0.068
\move(+1.867 +3.450)
\lcir r:+0.059
\move(+4.426 +0.867)
\lcir r:+0.075
\move(+5.295 +4.058)
\lcir r:+0.047
\move(+4.799 +1.586)
\lcir r:+0.111
\move(+4.803 +2.534)
\lcir r:+0.091
\move(+3.146 +2.976)
\lcir r:+0.063
\move(+1.306 +1.160)
\lcir r:+0.052
\move(+5.295 +0.095)
\lcir r:+0.055
\move(+4.058 +3.323)
\lcir r:+0.117
\move(+5.289 +2.202)
\lcir r:+0.110
\move(+2.853 +4.740)
\lcir r:+0.126
\move(+3.304 +1.962)
\lcir r:+0.102
\move(+5.325 +3.586)
\lcir r:+0.046
\move(+1.254 +4.450)
\lcir r:+0.074
\move(+4.103 +1.845)
\lcir r:+0.111
\move(+4.454 +3.545)
\lcir r:+0.077
\move(+3.501 +2.458)
\lcir r:+0.068
\move(+3.445 +3.609)
\lcir r:+0.047
\move(+3.581 +3.595)
\lcir r:+0.090
\move(+1.592 +5.298)
\lcir r:+0.108
\move(+3.980 +0.605)
\lcir r:+0.087
\move(+0.022 +0.394)
\lcir r:+0.137
\move(+3.748 +1.991)
\lcir r:+0.088
\move(+2.654 +1.259)
\lcir r:+0.054
\move(+2.584 +0.040)
\lcir r:+0.063
\move(+1.085 +2.308)
\lcir r:+0.110
\move(+2.631 +1.488)
\lcir r:+0.054
\move(+2.146 +1.660)
\lcir r:+0.047
\move(+1.070 +5.058)
\lcir r:+0.126
\move(+4.837 +3.996)
\lcir r:+0.079
\move(+1.731 +0.885)
\lcir r:+0.123
\move(+4.973 +2.606)
\lcir r:+0.095
\move(+1.738 +4.002)
\lcir r:+0.135
\move(+4.321 +2.393)
\lcir r:+0.074
\move(+3.255 +3.857)
\lcir r:+0.113
\move(+3.167 +2.114)
\lcir r:+0.103
\move(+1.373 +2.206)
\lcir r:+0.062
\move(+1.412 +3.818)
\lcir r:+0.046
\move(+3.928 +1.956)
\lcir r:+0.096
\move(+2.471 +4.138)
\lcir r:+0.048
\move(+3.720 +4.456)
\lcir r:+0.076
\move(+3.894 +4.663)
\lcir r:+0.056
\move(+3.018 +4.488)
\lcir r:+0.063
\move(+1.293 +3.441)
\lcir r:+0.048
\move(+4.411 +2.576)
\lcir r:+0.129
\move(+2.033 +1.340)
\lcir r:+0.076
\move(+3.009 +2.211)
\lcir r:+0.056
\move(+0.291 +1.482)
\lcir r:+0.047
\move(+4.113 +4.093)
\lcir r:+0.137
\move(+1.068 +0.588)
\lcir r:+0.120
\move(+5.188 +4.761)
\lcir r:+0.079
\move(+3.335 +4.350)
\lcir r:+0.129
\move(+0.800 +1.578)
\lcir r:+0.128
\move(+0.837 +3.650)
\lcir r:+0.065
\move(+0.057 +4.035)
\lcir r:+0.069
\move(+3.301 +0.859)
\lcir r:+0.064
\move(+4.819 +2.189)
\lcir r:+0.104
\move(+3.534 +0.700)
\lcir r:+0.076
\move(+4.499 +0.759)
\lcir r:+0.056
\move(+0.677 +4.667)
\lcir r:+0.058
\move(+1.423 +1.190)
\lcir r:+0.069
\move(+3.022 +2.345)
\lcir r:+0.053
\move(+0.471 +4.264)
\lcir r:+0.048
\move(+4.004 +4.653)
\lcir r:+0.054
\move(+0.356 +0.701)
\lcir r:+0.108
\move(+3.907 +3.710)
\lcir r:+0.128
\move(+1.906 +1.309)
\lcir r:+0.055
\move(+0.864 +3.509)
\lcir r:+0.062
\move(+1.243 +2.223)
\lcir r:+0.070
\move(+1.973 +3.309)
\lcir r:+0.118
\move(+0.748 +1.791)
\lcir r:+0.091
\move(+2.601 +4.796)
\lcir r:+0.133
\move(+1.718 +2.954)
\lcir r:+0.097
\move(+1.295 +2.426)
\lcir r:+0.132
\move(+3.000 +1.659)
\lcir r:+0.113
\move(+4.008 +0.417)
\lcir r:+0.103
\move(+2.032 +2.397)
\lcir r:+0.115
\move(+2.337 +2.833)
\lcir r:+0.086
\move(+3.506 +1.564)
\lcir r:+0.137
\move(+3.149 +0.979)
\lcir r:+0.129
\move(+4.120 +0.576)
\lcir r:+0.056
\move(+4.534 +4.762)
\lcir r:+0.070
\move(+0.406 +2.573)
\lcir r:+0.128
\move(+4.440 +4.364)
\lcir r:+0.137
\move(+0.621 +0.773)
\lcir r:+0.048
\move(+3.845 +0.503)
\lcir r:+0.081
\move(+0.194 +1.583)
\lcir r:+0.093
\move(+0.316 +4.718)
\lcir r:+0.076
\move(+2.186 +2.931)
\lcir r:+0.093
\move(+3.899 +4.801)
\lcir r:+0.082
\move(+5.165 +3.484)
\lcir r:+0.066
\move(+4.918 +4.869)
\lcir r:+0.053
\move(+3.159 +0.053)
\lcir r:+0.099
\move(+0.594 +5.101)
\lcir r:+0.080
\move(+0.853 +0.093)
\lcir r:+0.116
\move(+0.620 +2.308)
\lcir r:+0.087
\move(+0.050 +0.820)
\lcir r:+0.053
\move(+1.204 +1.491)
\lcir r:+0.074
\move(+2.351 +3.721)
\lcir r:+0.088
\move(+4.082 +1.502)
\lcir r:+0.061
\move(+3.230 +1.270)
\lcir r:+0.073
\move(+0.496 +2.132)
\lcir r:+0.128
\move(+3.389 +1.732)
\lcir r:+0.067
\move(+3.342 +1.108)
\lcir r:+0.104
\move(+2.571 +2.297)
\lcir r:+0.085
\move(+0.941 +4.493)
\lcir r:+0.098
\move(+2.144 +0.036)
\lcir r:+0.059
\move(+0.282 +3.290)
\lcir r:+0.047
\move(+1.657 +0.131)
\lcir r:+0.065
\move(+0.627 +2.701)
\lcir r:+0.127
\move(+3.193 +4.202)
\lcir r:+0.077
\move(+3.287 +0.514)
\lcir r:+0.109
\move(+1.775 +0.020)
\lcir r:+0.097
\move(+1.534 +5.006)
\lcir r:+0.084
\move(+1.522 +1.984)
\lcir r:+0.087
\move(+4.189 +2.848)
\lcir r:+0.092
\move(+5.177 +1.346)
\lcir r:+0.057
\move(+2.338 +4.885)
\lcir r:+0.063
\move(+2.368 +5.234)
\lcir r:+0.069
\move(+4.527 +5.190)
\lcir r:+0.068
\move(+1.163 +0.789)
\lcir r:+0.103
\move(+2.063 +4.295)
\lcir r:+0.074
\move(+0.004 +1.054)
\lcir r:+0.091
\move(+1.374 +0.786)
\lcir r:+0.108
\move(+1.630 +3.193)
\lcir r:+0.045
\move(+4.638 +0.361)
\lcir r:+0.110
\move(+4.659 +0.560)
\lcir r:+0.090
\move(+1.188 +4.849)
\lcir r:+0.114
\move(+2.296 +3.133)
\lcir r:+0.054
\move(+1.814 +4.507)
\lcir r:+0.076
\move(+5.247 +0.530)
\lcir r:+0.055
\move(+0.934 +0.488)
\lcir r:+0.047
\move(+0.544 +1.945)
\lcir r:+0.065
\move(+5.304 +3.076)
\lcir r:+0.044
\move(+4.509 +1.029)
\lcir r:+0.056
\move(+4.853 +4.191)
\lcir r:+0.048
\move(+0.252 +3.715)
\lcir r:+0.124
\move(+5.299 +1.618)
\lcir r:+0.053
\move(+0.685 +3.752)
\lcir r:+0.118
\move(+3.165 +0.720)
\lcir r:+0.130
\move(+4.053 +3.538)
\lcir r:+0.098
\move(+0.325 +4.336)
\lcir r:+0.115
\move(+0.221 +1.292)
\lcir r:+0.046
\move(+2.155 +5.016)
\lcir r:+0.129
\move(+5.238 +0.226)
\lcir r:+0.084
\move(+0.626 +0.126)
\lcir r:+0.114
\move(+4.771 +4.648)
\lcir r:+0.098
\move(+0.046 +1.486)
\lcir r:+0.075
\move(+2.320 +2.646)
\lcir r:+0.103
\move(+2.292 +1.232)
\lcir r:+0.057
\move(+2.842 +0.566)
\lcir r:+0.107
\move(+4.799 +0.704)
\lcir r:+0.111
\move(+0.238 +4.476)
\lcir r:+0.047
\move(+0.452 +5.125)
\lcir r:+0.064
\move(+4.198 +5.076)
\lcir r:+0.105
\move(+4.223 +1.667)
\lcir r:+0.104
\move(+3.068 +1.292)
\lcir r:+0.090
\move(+1.627 +0.528)
\lcir r:+0.069
\move(+2.597 +5.036)
\lcir r:+0.107
\move(+5.123 +2.247)
\lcir r:+0.061
\move(+2.159 +2.630)
\lcir r:+0.059
\move(+3.996 +0.760)
\lcir r:+0.069
\move(+0.163 +2.577)
\lcir r:+0.088
\move(+4.842 +2.345)
\lcir r:+0.053
\move(+3.539 +4.463)
\lcir r:+0.104
\move(+0.457 +5.361)
\lcir r:+0.102
\move(+2.718 +1.123)
\lcir r:+0.097
\move(+4.805 +0.194)
\lcir r:+0.126
\move(+0.099 +5.226)
\lcir r:+0.080
\move(+2.735 +2.205)
\lcir r:+0.097
\move(+5.143 +2.134)
\lcir r:+0.051
\move(+3.019 +3.773)
\lcir r:+0.138
\move(+3.362 +2.694)
\lcir r:+0.062
\move(+4.689 +3.932)
\lcir r:+0.061
\move(+1.890 +4.614)
\lcir r:+0.055
\move(+5.031 +3.048)
\lcir r:+0.093
\move(+2.640 +2.973)
\lcir r:+0.048
\move(+1.363 +1.006)
\lcir r:+0.113
\move(+3.048 +1.141)
\lcir r:+0.063
\move(+1.370 +3.918)
\lcir r:+0.053
\move(+3.705 +2.935)
\lcir r:+0.113
\move(+0.405 +1.951)
\lcir r:+0.074
\move(+4.655 +2.948)
\lcir r:+0.121
\move(+1.176 +3.507)
\lcir r:+0.051
\move(+4.910 +3.227)
\lcir r:+0.096
\move(+1.310 +3.599)
\lcir r:+0.111
\move(+0.892 +4.269)
\lcir r:+0.118
\move(+1.666 +3.780)
\lcir r:+0.098
\move(+3.714 +1.699)
\lcir r:+0.093
\move(+2.239 +1.935)
\lcir r:+0.105
\move(+4.848 +2.826)
\lcir r:+0.107
\move(+4.024 +1.653)
\lcir r:+0.096
\move(+4.540 +3.679)
\lcir r:+0.068
\move(+1.122 +1.129)
\lcir r:+0.135
\move(+4.953 +5.188)
\lcir r:+0.086
\move(+4.008 +4.322)
\lcir r:+0.116
\move(+0.491 +5.013)
\lcir r:+0.055
\move(+1.839 +3.587)
\lcir r:+0.080
\move(+0.874 +2.415)
\lcir r:+0.111
\move(+1.775 +4.711)
\lcir r:+0.096
\move(+3.446 +4.683)
\lcir r:+0.092
\move(+4.937 +4.310)
\lcir r:+0.088
\move(+4.609 +4.218)
\lcir r:+0.087
\move(+1.854 +2.515)
\lcir r:+0.079
\move(+0.710 +1.998)
\lcir r:+0.110
\move(+1.550 +1.676)
\lcir r:+0.078
\move(+3.081 +2.652)
\lcir r:+0.126
\move(+2.179 +0.277)
\lcir r:+0.053
\move(+0.369 +4.548)
\lcir r:+0.102
\move(+3.446 +0.831)
\lcir r:+0.083
\move(+1.498 +3.284)
\lcir r:+0.115
\move(+1.760 +5.211)
\lcir r:+0.082
\move(+0.860 +3.041)
\lcir r:+0.076
\move(+0.956 +3.598)
\lcir r:+0.065
\move(+3.142 +3.086)
\lcir r:+0.047
\move(+1.717 +0.667)
\lcir r:+0.096
\move(+4.545 +1.957)
\lcir r:+0.104
\move(+3.334 +3.281)
\lcir r:+0.074
\move(+2.821 +1.816)
\lcir r:+0.124
\move(+0.985 +4.667)
\lcir r:+0.077
\move(+0.874 +1.155)
\lcir r:+0.091
\move(+4.783 +3.884)
\lcir r:+0.045
\move(+5.009 +3.622)
\lcir r:+0.057
\move(+0.769 +4.429)
\lcir r:+0.085
\move(+0.741 +3.141)
\lcir r:+0.080
\move(+2.339 +4.034)
\lcir r:+0.085
\move(+1.584 +4.306)
\lcir r:+0.087
\move(+4.758 +2.401)
\lcir r:+0.049
\move(+4.732 +1.894)
\lcir r:+0.093
\move(+4.249 +3.321)
\lcir r:+0.074
\move(+0.156 +4.147)
\lcir r:+0.068
\move(+3.820 +1.826)
\lcir r:+0.073
\move(+1.024 +1.497)
\lcir r:+0.105
\move(+3.656 +0.212)
\lcir r:+0.073
\move(+1.708 +3.378)
\lcir r:+0.115
\move(+2.348 +3.004)
\lcir r:+0.085
\move(+4.505 +3.350)
\lcir r:+0.091
\move(+3.955 +1.172)
\lcir r:+0.104
\move(+0.987 +0.385)
\lcir r:+0.068
\move(+2.349 +0.884)
\lcir r:+0.090
\move(+0.536 +0.281)
\lcir r:+0.064
\move(+2.719 +0.920)
\lcir r:+0.107
\move(+4.657 +3.368)
\lcir r:+0.046
\move(+4.643 +1.397)
\lcir r:+0.090
\move(+4.474 +3.998)
\lcir r:+0.046
\move(+1.036 +4.369)
\lcir r:+0.058
\move(+0.006 +0.182)
\lcir r:+0.051
\move(+4.872 +3.868)
\lcir r:+0.046
\move(+3.389 +5.073)
\lcir r:+0.074
\move(+1.529 +0.890)
\lcir r:+0.079
\move(+0.931 +1.912)
\lcir r:+0.127
\move(+3.100 +5.051)
\lcir r:+0.122
\move(+0.301 +1.135)
\lcir r:+0.085
\move(+3.166 +3.404)
\lcir r:+0.135
\move(+2.894 +2.294)
\lcir r:+0.085
\move(+5.006 +1.904)
\lcir r:+0.093
\move(+1.897 +4.174)
\lcir r:+0.098
\move(+4.733 +4.116)
\lcir r:+0.062
\move(+5.010 +4.110)
\lcir r:+0.125
\move(+4.648 +2.173)
\lcir r:+0.068
\move(+2.228 +0.387)
\lcir r:+0.069
\move(+0.794 +4.593)
\lcir r:+0.080
\move(+1.107 +3.684)
\lcir r:+0.109
\move(+4.982 +2.747)
\lcir r:+0.047
\move(+3.482 +3.321)
\lcir r:+0.080
\move(+2.355 +0.137)
\lcir r:+0.057
\move(+2.916 +0.790)
\lcir r:+0.129
\move(+1.167 +4.319)
\lcir r:+0.083
\move(+0.176 +2.790)
\lcir r:+0.125
\move(+0.631 +3.222)
\lcir r:+0.044
\move(+2.272 +0.729)
\lcir r:+0.073
\move(+0.991 +1.696)
\lcir r:+0.096
\move(+3.754 +1.124)
\lcir r:+0.103
\move(+4.366 +2.035)
\lcir r:+0.091
\move(+5.152 +4.905)
\lcir r:+0.069
\move(+2.906 +1.392)
\lcir r:+0.101
\move(+1.205 +3.048)
\lcir r:+0.115
\move(+4.788 +5.219)
\lcir r:+0.082
\move(+2.607 +0.293)
\lcir r:+0.071
\move(+0.064 +1.338)
\lcir r:+0.072
\move(+4.343 +1.377)
\lcir r:+0.093
\move(+5.126 +0.422)
\lcir r:+0.107
\move(+0.785 +2.875)
\lcir r:+0.107
\move(+4.591 +4.530)
\lcir r:+0.069
\move(+4.052 +2.981)
\lcir r:+0.099
\move(+3.691 +1.288)
\lcir r:+0.073
\move(+3.534 +4.275)
\lcir r:+0.084
\move(+5.157 +3.911)
\lcir r:+0.122
\move(+1.382 +3.127)
\lcir r:+0.058
\move(+0.790 +3.339)
\lcir r:+0.124
\move(+3.632 +3.373)
\lcir r:+0.074
\move(+2.273 +4.321)
\lcir r:+0.072
\move(+0.261 +4.886)
\lcir r:+0.053
\move(+1.341 +4.775)
\lcir r:+0.056
\move(+4.322 +3.173)
\lcir r:+0.090
\move(+1.415 +4.063)
\lcir r:+0.095
\move(+4.831 +5.020)
\lcir r:+0.121
\move(+5.314 +2.638)
\lcir r:+0.129
\move(+1.490 +3.699)
\lcir r:+0.095
\move(+2.674 +0.583)
\lcir r:+0.062
\move(+2.128 +0.817)
\lcir r:+0.096
\move(+0.920 +0.953)
\lcir r:+0.116
\move(+2.413 +1.241)
\lcir r:+0.064
\move(+0.503 +4.866)
\lcir r:+0.077
\move(+4.611 +3.495)
\lcir r:+0.089
\move(+4.906 +4.563)
\lcir r:+0.061
\move(+2.269 +0.551)
\lcir r:+0.100
\move(+4.504 +2.137)
\lcir r:+0.081
\move(+1.234 +5.244)
\lcir r:+0.074
\move(+0.536 +4.386)
\lcir r:+0.085
\move(+1.050 +0.154)
\lcir r:+0.090
\move(+4.287 +3.010)
\lcir r:+0.075
\move(+2.895 +3.526)
\lcir r:+0.138
\move(+4.894 +5.360)
\lcir r:+0.096
\move(+1.291 +0.487)
\lcir r:+0.072
\move(+0.818 +0.625)
\lcir r:+0.133
\move(+1.394 +1.309)
\lcir r:+0.054
\move(+0.347 +3.361)
\lcir r:+0.048
\move(+3.912 +4.157)
\lcir r:+0.075
\move(+4.844 +3.049)
\lcir r:+0.093
\move(+2.456 +4.915)
\lcir r:+0.051
\move(+2.030 +2.829)
\lcir r:+0.093
\move(+0.109 +2.226)
\lcir r:+0.070
\move(+0.711 +4.291)
\lcir r:+0.065
\move(+3.079 +3.202)
\lcir r:+0.085
\move(+3.795 +4.286)
\lcir r:+0.100
\move(+0.918 +3.822)
\lcir r:+0.125
\move(+4.528 +4.899)
\lcir r:+0.064
\move(+0.436 +3.199)
\lcir r:+0.133
\move(+3.717 +4.713)
\lcir r:+0.092
\move(+4.142 +2.305)
\lcir r:+0.126
\move(+0.432 +4.762)
\lcir r:+0.048
\move(+0.677 +2.483)
\lcir r:+0.097
\move(+3.537 +3.063)
\lcir r:+0.097
\move(+3.431 +1.855)
\lcir r:+0.063
\move(+3.585 +0.840)
\lcir r:+0.057
\move(+4.618 +0.715)
\lcir r:+0.070
\move(+2.657 +0.165)
\lcir r:+0.053
\move(+1.092 +0.303)
\lcir r:+0.065
\move(+4.794 +1.414)
\lcir r:+0.058
\move(+2.431 +2.473)
\lcir r:+0.103
\move(+0.086 +3.008)
\lcir r:+0.111
\move(+3.944 +1.793)
\lcir r:+0.056
\move(+1.966 +0.968)
\lcir r:+0.126
\move(+1.724 +4.216)
\lcir r:+0.079
\move(+2.776 +3.065)
\lcir r:+0.117
\move(+4.874 +1.774)
\lcir r:+0.092
\move(+0.516 +4.518)
\lcir r:+0.048
\move(+5.145 +0.057)
\lcir r:+0.079
\move(+3.885 +3.143)
\lcir r:+0.133
\move(+1.730 +2.523)
\lcir r:+0.045
\move(+3.364 +1.273)
\lcir r:+0.062
\move(+1.967 +0.362)
\lcir r:+0.123
\move(+5.152 +2.435)
\lcir r:+0.129
\move(+0.015 +0.650)
\lcir r:+0.120
\move(+4.999 +0.783)
\lcir r:+0.097
\move(+5.294 +1.392)
\lcir r:+0.069
\move(+3.099 +3.987)
\lcir r:+0.091
\move(+2.909 +1.032)
\lcir r:+0.114
\move(+5.186 +3.064)
\lcir r:+0.062
\move(+3.255 +2.669)
\lcir r:+0.048
\move(+2.789 +4.152)
\lcir r:+0.088
\move(+2.718 +2.890)
\lcir r:+0.067
\move(+4.678 +3.644)
\lcir r:+0.074
\move(+4.566 +0.161)
\lcir r:+0.102
\move(+0.169 +1.878)
\lcir r:+0.105
\move(+1.272 +3.805)
\lcir r:+0.095
\move(+0.130 +1.187)
\lcir r:+0.093
\move(+0.037 +0.920)
\lcir r:+0.048
\move(+2.513 +0.925)
\lcir r:+0.057
\move(+0.364 +4.844)
\lcir r:+0.058
\move(+3.832 +1.286)
\lcir r:+0.064
\move(+0.209 +0.565)
\lcir r:+0.092
\move(+4.159 +1.367)
\lcir r:+0.090
\move(+0.269 +1.700)
\lcir r:+0.046
\move(+5.113 +5.096)
\lcir r:+0.098
\move(+1.618 +2.757)
\lcir r:+0.125
\move(+4.314 +3.561)
\lcir r:+0.064
\move(+1.521 +3.907)
\lcir r:+0.094
\move(+4.216 +4.314)
\lcir r:+0.092
\move(+4.910 +1.126)
\lcir r:+0.088
\move(+0.360 +1.797)
\lcir r:+0.087
\move(+2.017 +3.067)
\lcir r:+0.124
\move(+4.755 +4.236)
\lcir r:+0.060
\move(+0.051 +1.632)
\lcir r:+0.059
\move(+2.405 +2.254)
\lcir r:+0.085
\move(+0.914 +3.183)
\lcir r:+0.076
\move(+2.891 +2.135)
\lcir r:+0.074
\move(+2.319 +4.519)
\lcir r:+0.131
\move(+1.267 +1.282)
\lcir r:+0.076
\move(+4.248 +0.651)
\lcir r:+0.093
\move(+3.390 +3.474)
\lcir r:+0.099
\move(+1.719 +4.581)
\lcir r:+0.045
\move(+0.134 +4.295)
\lcir r:+0.081
\move(+0.685 +2.171)
\lcir r:+0.065
\move(+0.513 +0.691)
\lcir r:+0.050
\move(+2.540 +2.076)
\lcir r:+0.137
\move(+4.431 +0.315)
\lcir r:+0.103
\move(+3.762 +2.410)
\lcir r:+0.073
\move(+0.994 +4.866)
\lcir r:+0.080
\move(+4.168 +5.270)
\lcir r:+0.091
\move(+3.592 +2.104)
\lcir r:+0.054
\move(+5.034 +4.796)
\lcir r:+0.078
\move(+2.894 +0.409)
\lcir r:+0.058
\move(+2.660 +3.262)
\lcir r:+0.112
\move(+0.394 +1.335)
\lcir r:+0.133
\move(+4.262 +4.470)
\lcir r:+0.070
\move(+1.339 +4.355)
\lcir r:+0.053
\move(+4.387 +0.720)
\lcir r:+0.062
\move(+2.385 +5.060)
\lcir r:+0.106
\move(+2.728 +3.420)
\lcir r:+0.060
\move(+3.484 +0.170)
\lcir r:+0.104
\move(+2.154 +1.433)
\lcir r:+0.078
\move(+2.113 +1.562)
\lcir r:+0.057
\move(+2.620 +2.774)
\lcir r:+0.085
\move(+1.655 +5.100)
\lcir r:+0.070
\move(+5.056 +1.165)
\lcir r:+0.045
\move(+3.739 +3.799)
\lcir r:+0.061
\move(+3.870 +1.700)
\lcir r:+0.063
\move(+1.852 +2.729)
\lcir r:+0.111
\move(+1.885 +4.394)
\lcir r:+0.058
\move(+3.690 +3.113)
\lcir r:+0.064
\move(+0.527 +5.219)
\lcir r:+0.056
\move(+2.580 +2.482)
\lcir r:+0.046
\move(+5.298 +4.354)
\lcir r:+0.087
\move(+2.555 +3.746)
\lcir r:+0.098
\move(+5.355 +3.904)
\lcir r:+0.076
\move(+1.134 +2.690)
\lcir r:+0.122
\move(+2.236 +0.172)
\lcir r:+0.067
\move(+3.518 +0.503)
\lcir r:+0.122
\move(+4.419 +2.843)
\lcir r:+0.138
\move(+0.723 +3.552)
\lcir r:+0.086
\move(+3.330 +2.272)
\lcir r:+0.125
\move(+2.547 +1.175)
\lcir r:+0.081
\move(+1.247 +0.199)
\lcir r:+0.112
\move(+4.132 +4.716)
\lcir r:+0.090
\move(+4.836 +0.932)
\lcir r:+0.120
\move(+2.637 +2.574)
\lcir r:+0.062
\move(+1.163 +0.421)
\lcir r:+0.072
\move(+4.694 +4.838)
\lcir r:+0.107
\move(+2.208 +2.285)
\lcir r:+0.093
\move(+4.646 +2.694)
\lcir r:+0.134
\move(+2.771 +0.027)
\lcir r:+0.125
\move(+3.856 +3.346)
\lcir r:+0.072
\move(+1.026 +3.428)
\lcir r:+0.118
\move(+0.289 +4.160)
\lcir r:+0.065
\move(+3.957 +2.159)
\lcir r:+0.109
\move(+0.302 +3.003)
\lcir r:+0.104
\move(+0.621 +1.626)
\lcir r:+0.057
\move(+2.598 +4.087)
\lcir r:+0.048
\move(+2.507 +2.916)
\lcir r:+0.096
\move(+2.928 +1.236)
\lcir r:+0.054
\move(+1.863 +1.804)
\lcir r:+0.073
\move(+2.582 +3.908)
\lcir r:+0.066
\move(+1.459 +4.225)
\lcir r:+0.061
\move(+5.246 +2.883)
\lcir r:+0.125
\move(+3.650 +0.016)
\lcir r:+0.123
\move(+2.758 +5.170)
\lcir r:+0.102
\move(+2.394 +0.305)
\lcir r:+0.116
\move(+2.194 +5.248)
\lcir r:+0.106
\move(+3.053 +2.861)
\lcir r:+0.085
\move(+3.470 +3.939)
\lcir r:+0.118
\move(+1.794 +1.609)
\lcir r:+0.051
\move(+4.408 +5.014)
\lcir r:+0.101
\move(+3.004 +0.614)
\lcir r:+0.063
\move(+4.429 +4.613)
\lcir r:+0.113
\move(+4.299 +0.236)
\lcir r:+0.047
\move(+0.029 +3.708)
\lcir r:+0.091
\move(+0.101 +3.550)
\lcir r:+0.082
\move(+3.541 +3.459)
\lcir r:+0.051
\move(+2.205 +3.076)
\lcir r:+0.054
\move(+5.209 +2.038)
\lcir r:+0.065
\move(+2.993 +3.037)
\lcir r:+0.101
\move(+2.914 +5.099)
\lcir r:+0.070
\move(+0.138 +4.874)
\lcir r:+0.070
\move(+3.746 +5.187)
\lcir r:+0.098
\move(+3.702 +0.456)
\lcir r:+0.069
\move(+2.700 +0.441)
\lcir r:+0.081
\move(+0.414 +2.816)
\lcir r:+0.115
\move(+1.389 +5.300)
\lcir r:+0.090
\move(+3.707 +0.635)
\lcir r:+0.109
\move(+4.103 +0.691)
\lcir r:+0.057
\move(+0.292 +0.844)
\lcir r:+0.048
\move(+0.019 +3.237)
\lcir r:+0.128
\move(+3.557 +1.779)
\lcir r:+0.084
\move(+1.898 +5.275)
\lcir r:+0.070
\move(+4.069 +4.985)
\lcir r:+0.053
\move(+3.657 +1.421)
\lcir r:+0.065
\move(+1.137 +1.358)
\lcir r:+0.074
\move(+4.845 +2.671)
\lcir r:+0.047
\move(+3.158 +4.459)
\lcir r:+0.079
\move(+4.451 +2.395)
\lcir r:+0.055
\move(+5.193 +4.085)
\lcir r:+0.055
\move(+2.508 +2.645)
\lcir r:+0.085
\move(+2.031 +5.334)
\lcir r:+0.075
\move(+0.129 +1.720)
\lcir r:+0.059
\move(+2.503 +4.452)
\lcir r:+0.064
\move(+4.391 +1.559)
\lcir r:+0.096
\move(+4.872 +4.445)
\lcir r:+0.061
\move(+0.650 +0.287)
\lcir r:+0.045
\move(+2.186 +2.762)
\lcir r:+0.076
\move(+2.233 +2.470)
\lcir r:+0.093
\move(+1.647 +0.319)
\lcir r:+0.123
\move(+2.578 +3.085)
\lcir r:+0.079
\move(+4.196 +3.693)
\lcir r:+0.113
\move(+5.319 +3.448)
\lcir r:+0.092
\move(+4.782 +3.524)
\lcir r:+0.085
\move(+0.825 +0.399)
\lcir r:+0.093
\move(+3.163 +1.161)
\lcir r:+0.055
\move(+1.333 +1.951)
\lcir r:+0.106
\move(+3.323 +2.517)
\lcir r:+0.119
\move(+2.528 +0.168)
\lcir r:+0.076
\move(+4.688 +0.000)
\lcir r:+0.100
\move(+2.416 +0.760)
\lcir r:+0.052
\move(+4.362 +5.367)
\lcir r:+0.044
\move(+2.532 +1.361)
\lcir r:+0.106
\move(+1.627 +4.773)
\lcir r:+0.063
\move(+4.785 +3.359)
\lcir r:+0.081
\move(+0.040 +2.404)
\lcir r:+0.120
\move(+2.257 +1.558)
\lcir r:+0.084
\move(+5.280 +5.292)
\lcir r:+0.110
\move(+2.877 +2.882)
\lcir r:+0.092
\move(+0.758 +0.251)
\lcir r:+0.068
\move(+2.883 +4.349)
\lcir r:+0.131
\move(+0.882 +2.657)
\lcir r:+0.132
\move(+4.457 +3.215)
\lcir r:+0.051
\move(+2.458 +2.767)
\lcir r:+0.045
\move(+3.079 +1.471)
\lcir r:+0.090
\move(+1.984 +3.533)
\lcir r:+0.075
\move(+4.719 +2.052)
\lcir r:+0.065
\move(+4.682 +4.341)
\lcir r:+0.057
\move(+5.257 +5.133)
\lcir r:+0.051
\move(+0.523 +3.629)
\lcir r:+0.085
\move(+0.643 +4.534)
\lcir r:+0.079
\move(+1.444 +1.413)
\lcir r:+0.061
\move(+2.111 +2.537)
\lcir r:+0.046
\move(+5.157 +1.521)
\lcir r:+0.120
\move(+2.302 +3.310)
\lcir r:+0.048
\move(+3.450 +3.189)
\lcir r:+0.055
\move(+2.327 +1.386)
\lcir r:+0.101
\move(+0.537 +2.458)
\lcir r:+0.045
\move(+0.773 +1.336)
\lcir r:+0.116
\move(+4.356 +3.427)
\lcir r:+0.077
\move(+3.904 +0.940)
\lcir r:+0.134
\move(+4.284 +0.329)
\lcir r:+0.044
\move(+0.604 +3.100)
\lcir r:+0.062
\move(+3.565 +1.142)
\lcir r:+0.087
\move(+1.788 +0.488)
\lcir r:+0.097
\move(+1.245 +4.634)
\lcir r:+0.109
\move(+1.220 +2.863)
\lcir r:+0.071
\move(+4.614 +4.418)
\lcir r:+0.046
\move(+2.996 +5.273)
\lcir r:+0.123
\move(+2.050 +4.507)
\lcir r:+0.138
\move(+3.890 +2.927)
\lcir r:+0.071
\move(+5.337 +4.847)
\lcir r:+0.093
\move(+0.527 +4.660)
\lcir r:+0.091
\move(+5.186 +4.224)
\lcir r:+0.085
\move(+2.734 +2.421)
\lcir r:+0.119
\move(+3.419 +3.741)
\lcir r:+0.087
\move(+3.928 +2.370)
\lcir r:+0.097
\move(+4.390 +0.481)
\lcir r:+0.068
\move(+5.007 +0.580)
\lcir r:+0.091
\move(+4.305 +4.101)
\lcir r:+0.055
\move(+4.316 +3.910)
\lcir r:+0.136
\move(+2.688 +4.535)
\lcir r:+0.138
\move(+3.825 +2.816)
\lcir r:+0.056
\move(+3.126 +2.444)
\lcir r:+0.085
\move(+4.241 +1.499)
\lcir r:+0.066
\move(+2.440 +3.217)
\lcir r:+0.113
\move(+3.590 +4.623)
\lcir r:+0.064
\move(+3.803 +4.576)
\lcir r:+0.070
\move(+3.592 +3.773)
\lcir r:+0.088
\move(+3.251 +1.592)
\lcir r:+0.120
\move(+1.586 +1.086)
\lcir r:+0.125
\move(+0.257 +5.077)
\lcir r:+0.137
\move(+3.687 +1.853)
\lcir r:+0.063
\move(+1.490 +3.496)
\lcir r:+0.097
\move(+3.606 +3.225)
\lcir r:+0.077
\move(+0.930 +2.828)
\lcir r:+0.045
\move(+0.920 +0.266)
\lcir r:+0.069
\move(+4.224 +0.460)
\lcir r:+0.100
\move(+0.981 +0.770)
\lcir r:+0.074
\move(+3.415 +4.143)
\lcir r:+0.093
\move(+1.045 +5.308)
\lcir r:+0.126
\move(+1.318 +1.412)
\lcir r:+0.064
\move(+3.373 +0.954)
\lcir r:+0.054
\move(+4.079 +3.864)
\lcir r:+0.095
\move(+0.463 +3.843)
\lcir r:+0.122
\move(+3.788 +2.253)
\lcir r:+0.085
\move(+2.611 +1.872)
\lcir r:+0.063
\move(+2.496 +4.620)
\lcir r:+0.072
\move(+0.112 +3.834)
\lcir r:+0.060
\move(+1.475 +1.820)
\lcir r:+0.084
\move(+5.170 +0.646)
\lcir r:+0.085
\move(+5.180 +3.670)
\lcir r:+0.121
\move(+4.604 +4.048)
\lcir r:+0.083
\move(+5.015 +4.953)
\lcir r:+0.075
\move(+5.359 +4.185)
\lcir r:+0.093
\move(+3.868 +1.541)
\lcir r:+0.096
\move(+1.570 +4.459)
\lcir r:+0.067
\move(+1.740 +1.271)
\lcir r:+0.115
\move(+4.942 +2.434)
\lcir r:+0.081
\move(+1.085 +4.568)
\lcir r:+0.064
\move(+3.555 +0.322)
\lcir r:+0.064
\move(+2.427 +1.877)
\lcir r:+0.092
\move(+5.155 +1.225)
\lcir r:+0.066
\move(+3.569 +1.957)
\lcir r:+0.094
\move(+0.406 +0.876)
\lcir r:+0.069
\move(+3.294 +0.182)
\lcir r:+0.087
\move(+1.218 +4.017)
\lcir r:+0.107
\move(+2.015 +5.175)
\lcir r:+0.084
\move(+4.506 +0.623)
\lcir r:+0.075
\move(+2.112 +3.689)
\lcir r:+0.114
\move(+1.545 +3.065)
\lcir r:+0.109
\move(+1.079 +3.223)
\lcir r:+0.094
\move(+1.826 +0.202)
\lcir r:+0.091
\move(+1.713 +1.730)
\lcir r:+0.093
\move(+0.283 +2.227)
\lcir r:+0.104
\move(+5.105 +2.001)
\lcir r:+0.045
\move(+0.504 +1.107)
\lcir r:+0.120
\move(+2.889 +3.270)
\lcir r:+0.117
\move(+2.189 +1.767)
\lcir r:+0.068
\move(+3.792 +0.131)
\lcir r:+0.057
\move(+0.411 +1.597)
\lcir r:+0.119
\move(+3.740 +0.814)
\lcir r:+0.073
\move(+0.906 +4.988)
\lcir r:+0.052
\move(+1.449 +4.356)
\lcir r:+0.057
\move(+3.233 +4.649)
\lcir r:+0.125
\move(+0.081 +4.731)
\lcir r:+0.058
\move(+2.398 +4.735)
\lcir r:+0.079
\move(+0.402 +4.950)
\lcir r:+0.055
\move(+0.245 +3.970)
\lcir r:+0.130
\move(+1.538 +1.288)
\lcir r:+0.083
\move(+1.028 +4.131)
\lcir r:+0.076
\move(+1.154 +1.825)
\lcir r:+0.112
\move(+3.707 +1.541)
\lcir r:+0.065
\move(+2.666 +1.715)
\lcir r:+0.058
\move(+2.291 +2.124)
\lcir r:+0.088
\move(+1.012 +3.064)
\lcir r:+0.078
\move(+3.314 +3.065)
\lcir r:+0.126
\move(+1.737 +4.373)
\lcir r:+0.079
\move(+3.505 +0.985)
\lcir r:+0.082
\move(+5.309 +1.226)
\lcir r:+0.088
\move(+1.600 +2.139)
\lcir r:+0.087
\move(+4.488 +3.800)
\lcir r:+0.064
\move(+2.181 +4.750)
\lcir r:+0.138
\move(+5.264 +5.003)
\lcir r:+0.080
\move(+2.973 +0.126)
\lcir r:+0.101
\move(+2.572 +0.768)
\lcir r:+0.105
\move(+1.853 +1.462)
\lcir r:+0.107
\move(+3.749 +3.504)
\lcir r:+0.101
\move(+3.999 +2.778)
\lcir r:+0.111
\move(+4.045 +4.857)
\lcir r:+0.076
\move(+3.848 +2.706)
\lcir r:+0.057
\move(+4.391 +3.691)
\lcir r:+0.083
\move(+2.811 +2.665)
\lcir r:+0.134
\move(+1.128 +2.067)
\lcir r:+0.124
\move(+0.221 +3.174)
\lcir r:+0.084
\move(+4.074 +2.610)
\lcir r:+0.073
\move(+3.608 +4.107)
\lcir r:+0.100
\move(+1.731 +2.620)
\lcir r:+0.051
\move(+2.512 +3.425)
\lcir r:+0.107
\move(+4.356 +5.214)
\lcir r:+0.105
\move(+0.488 +0.788)
\lcir r:+0.051
\move(+1.270 +0.360)
\lcir r:+0.051
\move(+0.003 +1.995)
\lcir r:+0.095
\move(+1.573 +2.525)
\lcir r:+0.112
\move(+3.674 +3.933)
\lcir r:+0.087
\move(+4.768 +1.251)
\lcir r:+0.102
\move(+2.392 +1.074)
\lcir r:+0.104
\move(+3.383 +0.337)
\lcir r:+0.092
\move(+1.584 +4.620)
\lcir r:+0.095
\move(+1.377 +0.387)
\lcir r:+0.059
\move(+0.384 +0.337)
\lcir r:+0.098
\move(+4.379 +0.127)
\lcir r:+0.085
\move(+4.984 +2.277)
\lcir r:+0.081
\move(+2.091 +0.630)
\lcir r:+0.094
\move(+5.036 +2.869)
\lcir r:+0.085
\move(+1.473 +5.155)
\lcir r:+0.077
\move(+1.891 +3.788)
\lcir r:+0.127
\move(+2.433 +0.011)
\lcir r:+0.091
\move(+1.528 +0.656)
\lcir r:+0.093
\move(+3.148 +1.726)
\lcir r:+0.049
\move(+0.191 +4.760)
\lcir r:+0.056
\move(+4.649 +5.217)
\lcir r:+0.057
\move(+1.572 +4.139)
\lcir r:+0.080
\move(+3.970 +0.146)
\lcir r:+0.115
\move(+3.069 +0.443)
\lcir r:+0.120
\move(+1.481 +4.834)
\lcir r:+0.096
\move(+4.772 +0.459)
\lcir r:+0.057
\move(+0.746 +0.839)
\lcir r:+0.093
\move(+1.920 +1.174)
\lcir r:+0.081
\move(+4.796 +4.347)
\lcir r:+0.058
\move(+3.931 +4.953)
\lcir r:+0.074
\move(+3.444 +2.091)
\lcir r:+0.089
\move(+0.623 +0.661)
\lcir r:+0.064
\move(+1.394 +2.906)
\lcir r:+0.110
\move(+5.114 +2.718)
\lcir r:+0.085
\move(+4.211 +4.880)
\lcir r:+0.092
\move(+4.148 +0.077)
\lcir r:+0.075
\move(+3.775 +4.120)
\lcir r:+0.068
\move(+3.636 +2.482)
\lcir r:+0.069
\move(+2.764 +3.806)
\lcir r:+0.119
\move(+3.019 +4.162)
\lcir r:+0.101
\move(+2.721 +1.380)
\lcir r:+0.084
\move(+0.558 +3.424)
\lcir r:+0.124
\move(+0.552 +1.774)
\lcir r:+0.107
\move(+1.863 +5.027)
\lcir r:+0.129
\move(+3.859 +0.737)
\lcir r:+0.068
\move(+1.119 +4.451)
\lcir r:+0.058
\move(+3.556 +4.874)
\lcir r:+0.129
\move(+1.159 +4.175)
\lcir r:+0.061
\move(+0.373 +0.515)
\lcir r:+0.080
\move(+4.970 +1.533)
\lcir r:+0.068
\move(+3.670 +0.956)
\lcir r:+0.085
\move(+0.606 +0.465)
\lcir r:+0.133
\move(+4.498 +5.349)
\lcir r:+0.091
\move(+4.382 +0.600)
\lcir r:+0.051
\move(+3.702 +2.731)
\lcir r:+0.091
\move(+4.608 +0.896)
\lcir r:+0.110
\move(+1.094 +0.936)
\lcir r:+0.059
\move(+2.783 +1.575)
\lcir r:+0.120
\move(+0.884 +2.169)
\lcir r:+0.135
\move(+2.258 +3.625)
\lcir r:+0.046
\move(+4.612 +2.460)
\lcir r:+0.103
\move(+1.105 +4.708)
\lcir r:+0.050
\move(+0.504 +2.993)
\lcir r:+0.084
\move(+2.702 +4.001)
\lcir r:+0.087
\move(+0.577 +4.963)
\lcir r:+0.045
\move(+0.442 +2.351)
\lcir r:+0.097
\move(+1.147 +1.628)
\lcir r:+0.074
\move(+1.678 +4.914)
\lcir r:+0.087
\move(+1.254 +5.007)
\lcir r:+0.057
\move(+2.898 +1.993)
\lcir r:+0.068
\move(+5.325 +1.787)
\lcir r:+0.117
\move(+4.128 +0.831)
\lcir r:+0.081
\move(+2.691 +3.616)
\lcir r:+0.085
\move(+3.716 +3.673)
\lcir r:+0.066
\move(+1.872 +4.836)
\lcir r:+0.062
\move(+3.797 +5.021)
\lcir r:+0.076
\move(+3.659 +0.351)
\lcir r:+0.044
\move(+1.056 +4.251)
\lcir r:+0.047
\move(+2.197 +1.317)
\lcir r:+0.047
\move(+0.837 +4.771)
\lcir r:+0.104
\move(+4.163 +1.151)
\lcir r:+0.073
\move(+0.180 +0.776)
\lcir r:+0.084
\move(+4.770 +1.094)
\lcir r:+0.055
\move(+1.944 +0.554)
\lcir r:+0.070
\move(+4.422 +1.180)
\lcir r:+0.119
\move(+4.026 +5.341)
\lcir r:+0.068
\move(+1.393 +0.576)
\lcir r:+0.063
\move(+3.883 +3.961)
\lcir r:+0.124
\move(+4.476 +3.069)
\lcir r:+0.096
\move(+3.933 +4.552)
\lcir r:+0.061
\move(+0.706 +4.150)
\lcir r:+0.076
\move(+3.529 +5.259)
\lcir r:+0.053
\move(+0.657 +3.974)
\lcir r:+0.106
\move(+0.593 +1.279)
\lcir r:+0.074
\move(+2.582 +1.033)
\lcir r:+0.064
\move(+0.181 +3.390)
\lcir r:+0.095
\move(+4.350 +2.225)
\lcir r:+0.097
\move(+1.277 +3.258)
\lcir r:+0.107
\move(+0.284 +1.986)
\lcir r:+0.053
\move(+3.387 +4.538)
\lcir r:+0.065
\move(+0.368 +3.482)
\lcir r:+0.075
\move(+5.043 +0.997)
\lcir r:+0.097
\move(+4.612 +4.658)
\lcir r:+0.060
\move(+5.283 +4.526)
\lcir r:+0.085
\move(+4.988 +4.444)
\lcir r:+0.055
\move(+0.240 +0.233)
\lcir r:+0.078
\move(+0.464 +4.094)
\lcir r:+0.122
\move(+2.568 +5.243)
\lcir r:+0.102
\move(+4.221 +3.460)
\lcir r:+0.061
\move(+1.344 +1.658)
\lcir r:+0.125
\move(+4.977 +3.440)
\lcir r:+0.128
\move(+3.346 +1.415)
\lcir r:+0.082
\move(+1.945 +4.309)
\lcir r:+0.045
\move(+1.750 +1.873)
\lcir r:+0.054
\move(+3.258 +1.789)
\lcir r:+0.077
\move(+1.223 +0.018)
\lcir r:+0.070
\move(+4.151 +0.268)
\lcir r:+0.103
\move(+4.385 +4.823)
\lcir r:+0.091
\move(+2.075 +0.173)
\lcir r:+0.094
\move(+3.209 +1.399)
\lcir r:+0.056
\move(+2.549 +3.588)
\lcir r:+0.060
\move(+1.497 +0.457)
\lcir r:+0.080
\move(+2.782 +2.053)
\lcir r:+0.062
\move(+2.162 +4.362)
\lcir r:+0.046
\move(+2.544 +0.425)
\lcir r:+0.076
\move(+5.318 +4.681)
\lcir r:+0.074
\move(+1.935 +0.079)
\lcir r:+0.074
\move(+3.524 +2.635)
\lcir r:+0.111
\move(+4.148 +3.140)
\lcir r:+0.087
\move(+0.643 +4.799)
\lcir r:+0.078
\move(+2.799 +4.966)
\lcir r:+0.106
\move(+2.022 +1.483)
\lcir r:+0.064
\move(+2.425 +3.877)
\lcir r:+0.085
\move(+3.908 +3.468)
\lcir r:+0.062
\move(+1.485 +0.116)
\lcir r:+0.107
\move(+0.055 +4.605)
\lcir r:+0.048
\move(+0.717 +1.039)
\lcir r:+0.104
\move(+1.913 +0.736)
\lcir r:+0.112
\move(+3.253 +3.633)
\lcir r:+0.111
\move(+1.337 +5.160)
\lcir r:+0.059
\move(+2.367 +2.014)
\lcir r:+0.046
\move(+0.189 +4.620)
\lcir r:+0.084
\move(+2.808 +1.260)
\lcir r:+0.063
\move(+1.348 +4.914)
\lcir r:+0.059
\move(+0.586 +4.240)
\lcir r:+0.069
\move(+1.051 +2.500)
\lcir r:+0.085
\move(+0.250 +2.419)
\lcir r:+0.091
\move(+3.074 +4.821)
\lcir r:+0.109
\move(+2.900 +4.550)
\lcir r:+0.071
\move(+3.075 +1.895)
\lcir r:+0.134
\move(+0.107 +0.220)
\lcir r:+0.056
\move(+5.195 +1.086)
\lcir r:+0.080
\move(+3.738 +3.275)
\lcir r:+0.065
\move(+3.563 +5.106)
\lcir r:+0.103
\move(+2.157 +4.031)
\lcir r:+0.067
\move(+0.979 +1.305)
\lcir r:+0.092
\move(+2.735 +0.707)
\lcir r:+0.070
\move(+4.688 +2.305)
\lcir r:+0.070
\move(+5.174 +1.903)
\lcir r:+0.074
\move(+3.760 +4.874)
\lcir r:+0.075
\move(+1.662 +3.585)
\lcir r:+0.097
\move(+2.122 +0.472)
\lcir r:+0.067
\move(+5.111 +1.738)
\lcir r:+0.103
\move(+4.315 +1.850)
\lcir r:+0.101
\move(+2.068 +2.104)
\lcir r:+0.135
\move(+3.133 +2.286)
\lcir r:+0.073
\move(+4.731 +4.471)
\lcir r:+0.083
\move(+4.393 +1.001)
\lcir r:+0.063
\move(+2.004 +4.878)
\lcir r:+0.076
\move(+4.970 +1.659)
\lcir r:+0.058
\move(+4.202 +2.512)
\lcir r:+0.089
\move(+0.866 +5.292)
\lcir r:+0.055
\move(+0.060 +4.997)
\lcir r:+0.076
\move(+2.069 +4.144)
\lcir r:+0.077
\move(+1.382 +5.039)
\lcir r:+0.070
\move(+0.095 +4.467)
\lcir r:+0.096
\move(+2.899 +3.989)
\lcir r:+0.109
\move(+3.267 +5.185)
\lcir r:+0.092
\move(+2.944 +2.475)
\lcir r:+0.098
\move(+4.275 +4.727)
\lcir r:+0.054
\move(+4.929 +4.690)
\lcir r:+0.066
\move(+4.962 +3.928)
\lcir r:+0.062
\move(+2.502 +0.586)
\lcir r:+0.090
\move(+0.476 +0.596)
\lcir r:+0.052
\move(+2.750 +5.938)
\rlvec(+0.000 -0.100)
\rmove(+0.000 +0.050)
\rlvec(+0.992 +0.000)
\rmove(+0.000 +0.050)
\rlvec(+0.000 -0.050)
\rlvec(+0.992 +0.000)
\rmove(+0.000 +0.050)
\rlvec(+0.000 -0.100)
\move(+2.750 +5.938)
\Ttext{ 0}
\move(+4.613 +5.978)
\htext{$\mathsf{10\,D_{50}}$}
\move(+3.742 +5.938)
\Btext{Length scale}
\end{texdraw}
\normalsize\rmfamily

%% file: McNu96_paper_arxiv.bbl
\begin{thebibliography}{}

\bibitem[\protect\citeauthoryear{}{Alzebdeh and
  Ostoja-Starzewski}{1995}]{Ostoja:1995a}
Alzebdeh, K. and Ostoja-Starzewski, M. (1995).
\newblock ``On the effective elastic moduli of granular materials.''\ {\em
  Engineering Mechanics: Proc. of the 10th Conf.}, S. Sture, ed., Vol.~1, ASCE,
  New York, N.Y.,  639--641.

\bibitem[\protect\citeauthoryear{}{Annic et~al.\@}{1993}]{Annic:1993a}
Annic, C., Bideau, D., Lema{\^{i}}tre, J., Troadec, J.~P., and Gervois, A.
  (1993).
\newblock ``Geometrical properties of {2D} packings of particles.''\ {\em
  Powders \& Grains 93}, C. Thornton, ed., A.A. Balkema, Rotterdam, The
  Netherlands,  11--16.

\bibitem[\protect\citeauthoryear{}{Bagi}{1996}]{Bagi:1996a}
Bagi, K. (1996).
\newblock ``Stress and strain in granular assemblies.''\ {\em Mech. of Mater.},
  22(3), 165--177.

\bibitem[\protect\citeauthoryear{}{Bardet and Proubet}{1991}]{Bardet:1991a}
Bardet, J.~P. and Proubet, J. (1991).
\newblock ``A numerical investigation of the structure of persistent shear
  bands in granular media.''\ {\em G{\'{e}}otechnique}, 41(4), 599--613.

\bibitem[\protect\citeauthoryear{}{Christoffersen
  et~al.\@}{1981}]{Christoffersen:1981a}
Christoffersen, J., Mehrabadi, M.~M., and Nemat-Nasser, S. (1981).
\newblock ``A micromechanical description of granular material behavior.''\
  {\em J. Appl. Mech.}, ASME, 48(2), 339--344.

\bibitem[\protect\citeauthoryear{}{Cundall et~al.\@}{1982}]{Cundall:1982a}
Cundall, P.~A., Drescher, A., and Strack, O. D.~L. (1982).
\newblock ``Numerical experiments on granular assemblies: measurements and
  observations.''\ {\em Deformation and Failure of Granular Materials}, P.
  Vermeer and H. Luger, eds., A.A. Balkema, Rotterdam, the Netherlands,
  355--370.

\bibitem[\protect\citeauthoryear{}{Drescher and de~Josselin~de
  Jong}{1972}]{Drescher:1972a}
Drescher, A. and de~Josselin~de Jong, G. (1972).
\newblock ``Photoelastic verification of a mechanical model for the flow of a
  granular material.''\ {\em J. Mech. Phys. Solids}, 20, 337--351.

\bibitem[\protect\citeauthoryear{}{Horne}{1965}]{Horne:1965b}
Horne, M.~R. (1965).
\newblock ``The behaviour of an assembly of rotund, rigid, cohesionless
  particles, {II}.''\ {\em Proc. R. Soc. Lond. A}, 286, 79--97.

\bibitem[\protect\citeauthoryear{}{Knuth}{1973}]{Knuth:1973a}
Knuth, D.~E. (1973).
\newblock {\em The Art of Computer Programming: Fundamental Algorithms},
  Vol.~1.
\newblock Addison-Wesley Pub., Reading, Mass.

\bibitem[\protect\citeauthoryear{}{Kuhn}{1996}]{Kuhn:1996a}
Kuhn, M.~R. (1996).
\newblock ``Experimental measurement of strain gradient effects in granular
  materials.''\ {\em Engineering Mechanics: Proc. of the 11th Conference},
  Y.~K. Lin and T.~C. Su, eds., Vol.~2, ASCE, New York, N.Y.,  881--885.

\bibitem[\protect\citeauthoryear{}{Love}{1927}]{Love:1927a}
Love, A. E.~H. (1927).
\newblock {\em A Treatise on the Mathematical Theory of Elasticity}.
\newblock Dover Pub., New York, N.Y., 4th edition.

\bibitem[\protect\citeauthoryear{}{Oda}{1975}]{Oda:1975a}
Oda, M. (1975).
\newblock ``On stress-dilatancy relation of sand in simple shear test.''\ {\em
  Soils and Found.}, Jap. Soc. Soil Mech. Found. Eng., 15(2), 17--29.

\bibitem[\protect\citeauthoryear{}{Oda et~al.\@}{1982}]{Oda:1982a}
Oda, M., Konishi, J., and Nemat-Nasser, S. (1982).
\newblock ``Experimental micromechanical evaluation of strength of granular
  materials: effects of particle rolling.''\ {\em Mech. of Mater.}, 1(4),
  269--283.

\bibitem[\protect\citeauthoryear{}{O'Rourke}{1994}]{Orourke:1994a}
O'Rourke, J. (1994).
\newblock {\em Computational Geometry in {C}}.
\newblock Cambridge University Press, Cambridge, U.K.

\bibitem[\protect\citeauthoryear{}{Preparata and
  Shamos}{1985}]{Preparata:1985a}
Preparata, F.~P. and Shamos, M.~I. (1985).
\newblock {\em Computational Geometry: An Introduction}.
\newblock Springer-Verlag, New York.

\bibitem[\protect\citeauthoryear{}{Rothenburg and
  Bathurst}{1992}]{Rothenburg:1992a}
Rothenburg, L. and Bathurst, R. (1992).
\newblock ``Effects of particle shape on micromechanical behavior of granular
  materials.''\ {\em Advances in micromechanics of granular materials}, H.
  Shen, M. Satake, M. Mehrabadi, C. Chang, and C. Campbell, eds., Elsevier,
  Amsterdam, The Netherlands,  343--352.

\bibitem[\protect\citeauthoryear{}{Rowe}{1962}]{Rowe:1962a}
Rowe, P.~W. (1962).
\newblock ``The stress-dilatancy relation for static equilibrium of an assembly
  of particles in contact.''\ {\em Proc. R. Soc. Lond. A}, London, 269,
  500--527.

\bibitem[\protect\citeauthoryear{}{Satake}{1993}]{Satake:1993a}
Satake, M. (1993).
\newblock ``Discrete-mechanical approach to granular media.''\ {\em Powders \&
  Grains 93}, C. Thornton, ed., A.A. Balkema, Rotterdam,  39--43.

\bibitem[\protect\citeauthoryear{}{Williams and Rege}{1997}]{Williams}
Williams, J.~R. and Rege, N. (1997).
\newblock ``Coherent vortex structures in deforming granular materials.''\ {\em
  Mechanics of Cohesive-frictional Materials}, 2(3), 223--236.

\end{thebibliography}
